\documentclass[aps,reprint,prb,twocolumn]{revtex4-1}

\usepackage{amsmath}
\usepackage{amsfonts}
\usepackage{amssymb}
\usepackage{graphicx}
\usepackage{moreverb}
\usepackage{url}
\usepackage{epsf}
\usepackage{color}
\usepackage{multirow}
\usepackage{subfigure}

\begin{document}

\title{System-size convergence of point defect properties: The case of the silicon vacancy}
\author{Fabiano Corsetti}
\email[E-mail: ]{fabiano.corsetti08@imperial.ac.uk}
\author{Arash A. Mostofi}
\affiliation{Thomas Young Centre, Imperial College London, London SW7 2AZ, United Kingdom}
\date{\today}

\begin{abstract}
We present a comprehensive study of the vacancy in bulk silicon in all its charge states from $2+$ to $2-$, using a supercell approach within plane-wave density-functional theory, and systematically quantify the various contributions to the well-known finite size errors associated with calculating formation energies and stable charge state transition levels of isolated defects with periodic boundary conditions. Furthermore, we find that transition levels converge faster with respect to supercell size when only the $\Gamma$-point is sampled in the Brillouin zone, as opposed to a dense k-point sampling. This arises from the fact that defect level at the $\Gamma$-point quickly converges to a fixed value which correctly describes the bonding at the defect centre. Our calculated transition levels with 1000-atom supercells and $\Gamma$-point only sampling are in good agreement with available experimental results. We also demonstrate two simple and accurate approaches for calculating the valence band offsets that are required for computing formation energies of charged defects, one based on a potential averaging scheme and the other using maximally-localized Wannier functions (MLWFs). Finally, we show that MLWFs provide a clear description of the nature of the electronic bonding at the defect centre that verifies the canonical Watkins model.
\end{abstract}

\maketitle

\section{Introduction}
\label{sec:intro}

The presence of point defects can have many consequences for the optical, electrical and mechanical properties of materials. In particular, the behaviour of defects in semiconductors has been the subject of thorough and ongoing research due to their use in devices such as transistors, solar cells and light-emitting diodes. This is because point defects strongly influence the electrical conductivity of semiconductors by adding states in the band gap, thus changing the number of charge carriers available. Therefore, both native defects (vacancies and self-interstitials) and impurity-related defects (dopants) play a crucial role in understanding and controlling the performance of semiconducting materials such as silicon, germanium and gallium arsenide.

To this end, the study of point defects in semiconductors using first-principles electronic structure simulations has received a considerable amount of attention in recent years.\cite{*[{For a review, see: }]defect-rev} Most notably, density-functional theory\cite{ks} (DFT) has proven extremely popular due to its balance of computational speed and predictive success. Nevertheless, it has been noted that even for one of the simplest cases of point defects in semiconductors, that of the neutral vacancy in silicon, theoretical studies have shown a large scatter in results for basic quantities such as the defect formation energy.\cite{niem-multisymm,probert}

For a point defect in an infinite crystal lattice the relaxation of atomic positions around the defect can extend to many successive shells of atoms. In simulations with period boundary conditions this gives rise to elastic interactions between the defect and its images in neighbouring supercells; the relaxation must therefore be contained within one supercell. Additionally, there is a spurious electrostatic interaction between a defect and its periodic replicas that depends on the size of the supercell.

We focus on the isolated vacancy in bulk silicon (denoted $\mathrm{v}^q$, where $q$ is the charge state of the defect centre). A number of previous studies have been undertaken on this defect centre using a variety of theoretical techniques;\footnote{For a review, see Probert and Payne\cite{probert} and Puska {\em et al.}\cite{niem-vac}.} in particular, Probert and Payne\cite{probert}, Puska {\em et al.}\cite{niem-vac}, and Wright\cite{wright-vac} have performed studies of the convergence of the defect formation energy.

In this Article we investigate systematically and quantify the main contributions to the finite size error that lead to the well-known slow convergence with respect to system size of, for example, the vacancy formation energy, which arise from spurious electrostatic interactions, elastic interactions and wavefunction orthogonality constraints between periodic images of the defect centre in the supercell approach. In addition, our calculations demonstrate that the defect formation energy and equilibrium charge state transition levels exhibit different convergence behaviour with respect to supercell size, depending on the Brillouin zone sampling used: the former benefits from the use of a dense k-point grid, while the latter from sampling at the $\Gamma$-point only. The reasons for this difference will be discussed in the text. Furthermore, we present two simple and accurate methods for calculating the potential alignment correction to the valence band maximum of charged defect supercells, one based on averaging the potential with a Voronoi cell construction, the other on matrix elements between maximally-localized Wannier functions\cite{mlwf} (MLWFs). Finally, we relate the MLWFs associated with the defect centre for each charge state to the canonical model of Watkins\cite{watkins-model,negU-baraff,vac} (described in Fig.~\ref{fig:split}, Appendix~\ref{app:res-mlwfs}) and show that the qualitative description given by this simple model is in full agreement with parameter-free DFT calculations.

The rest of the paper is organized as follows: in Sec.~\ref{sec:comp} we describe the computational techniques that we employ and give the technical details of our simulations. In Sec.~\ref{sec:pot_align} we illustrate the two methods we use to perform the potential alignment correction and show preliminary results comparing them. In Sec.~\ref{sec:results} we present our main results; first we describe the convergence properties of the unrelaxed (Sec.~\ref{subsec:neutral_unrel}) and relaxed (Sec.~\ref{subsec:neutral_rel}) neutral vacancy, and then we describe the results obtained for different charge states of the defect (Sec.~\ref{subsec:charged}) in terms of the transition levels (Sec.~\ref{subsec:trans_levels}). In Sec.~\ref{sec:conclusions} we give a brief summary of our main conclusions.

\section{Computational methods}
\label{sec:comp}

In the supercell approximation, the formation energy of the vacancy $E_f^q$ is defined (following Zhang and Northrup\cite{zhang}) as
\begin{equation}\label{eqn:E_f}
E_f^q = E_\mathrm{vac}^q - \left ( \frac{N-1}{N} \right ) E_\mathrm{bulk} + q \mu_e,
\end{equation}\\
where $E_\mathrm{bulk}$ is the total energy of the bulk (perfect crystal) supercell, $E_\mathrm{vac}^q$ is the total energy of the same supercell containing a single vacancy with charge $q$, $N$ is the number of atoms in the bulk supercell, and $\mu_e$ is the electron chemical potential. This last term can be divided into $\varepsilon_v + \Delta \mu_e$, where $\varepsilon_v$ is the value of the valence band maximum (VBM) and $\Delta \mu_e$ is the position of the Fermi level in the band gap. The determination of $\varepsilon_v$ is discussed in more detail in Sec.~\ref{sec:pot_align}.

For supercell sizes up to and including 256 atoms, the calculations are performed using the {\sc castep}\cite{castep} code (version 5.0). The Ceperley-Alder local-density approximation\cite{qmc1} (LDA) is used to describe exchange and correlation. For charged supercell calculations a compensating uniform background charge is added. We use two pseudopotentials for silicon: {\sc castep}'s `on-the-fly' Vanderbilt ultrasoft pseudopotential\cite{ultra-pseudo}, and a norm-conserving pseudopotential\cite{norm-pseudo}; both have four valence electrons and give accurate lattice constants for bulk silicon (5.39~$\mathrm{\AA}$ and 5.38~$\mathrm{\AA}$ respectively, compared with an experimental value of 5.43~$\mathrm{\AA}$). This slight underestimation of less than $1\%$ is typical for LDA DFT. The DFT-optimized lattice parameter is used in all calculations.

For our largest calculations, on a 1000-atom simple cubic (SC) supercell, we use the linear-scaling DFT code {\sc onetep}\cite{onetep1,onetep-nick} (version 2.4). The `cross-over', namely, the system-size at which it becomes computationally more efficient to use {\sc onetep} as opposed to conventional cubic-scaling DFT, is highly system dependent and lies at around 500 atoms for silicon. For this reason we use conventional plane-wave DFT for supercells smaller than this cross-over, and linear-scaling DFT for the largest supercell. {\sc onetep} makes use of the single-particle density matrix, expressed in separable form in terms of a localized basis of non-orthogonal generalized Wannier functions (NGWFs) and a density kernel.\cite{onetep1,onetep-nick} We use the same norm-conserving pseudopotential as for {\sc castep}, and nine NGWFs on each silicon atom with a truncation radius of 3.97~$\mathrm{\AA}$. We do not truncate the density kernel.

We follow the methodology outlined by Probert and Payne in their study of the neutral silicon vacancy\cite{probert}; however, numerical parameters are converged independently for all system sizes. The supercells are constructed from three unit cell shapes: FCC (the primitive cell, with 2 atoms), SC (with 8 atoms), and BCC (with 32 atoms). The supercells are then made from $n^3$ unit cells. We perform calculations on both unrelaxed and relaxed geometries; our convergence tolerance for the defect formation energy of a given supercell is 10~meV. We use a regular Monkhorst-Pack (MP) mesh\cite{mp_grid} of k-points for the Brillouin zone integration. In the text, the parameter $k_\mathrm{MP}$ refers to a $k_\mathrm{MP}^3$ grid. The issue of Brillouin zone sampling shall be discussed in Sec.~\ref{sec:results}; in general, we make use of two sampling schemes: a $\Gamma$-point only sampling (i.e., $k_\mathrm{MP} = 1$), and what we call a `dense' sampling. This last term we shall use to indicate that $k_\mathrm{MP}$ has been converged with respect to the formation energy for a particular supercell.

For the relaxation a quasi-Newton BFGS scheme is used. All the atoms in the supercell are allowed to move, and their positions are slightly randomized at the start of the procedure to allow for symmetry breaking. Our convergence tolerance for the geometry optimization is $5 \times 10^{-3}$~eV/$\mathrm{\AA}$ for the root mean square force on all the atoms.

\begin{figure}
  \includegraphics[width=0.4\textwidth]{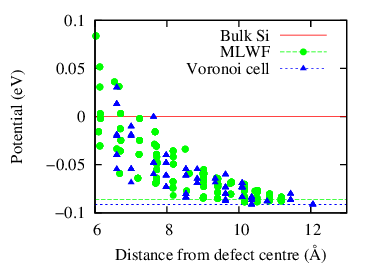}
  \caption{(Color online) Potential alignment correction for $\mathrm{v}^{2-}$ using the MLWF and real-space Voronoi cell methods. The values are adjusted with respect to the bulk supercell. The dashed horizontal lines show the value for the furthest point from the defect centre.\label{fig:pot_align}}
\end{figure}

\section{Determining the electron chemical potential}
\label{sec:pot_align}

As described in Sec.~\ref{sec:comp}, the formation energy of charged defects depends on the electron chemical potential, which is given relative to the VBM eigenvalue position $\varepsilon_v$. $\varepsilon_v$ is more precisely defined as the energy difference between the pure host material and the host with a single electron hole. In the limit of a dilute hole gas, this energy difference is equivalent to the VBM eigenvalue. For a finite supercell calculation, however, the energy difference between the bulk system with and without the hole only slowly converges to the infinite limit as the system size is increased.\cite{vbm_hole} Instead, the VBM will always remain at the same position (given a sufficient Brillouin zone sampling), which is the limiting value.

Two approaches can be taken to determine $\varepsilon_v$: either calculating the energy difference of the bulk supercell with $q = 0$ and $q = {+1}$, or using the bulk VBM eigenvalue and employ a potential alignment correction\cite{vbm_hole,def_corr,niem-gaas}. In the rest of this section we describe the two methods that we employ for calculating the potential alignment correction: a partitioning of the real-space potential using Voronoi cells, and an MLWF-based approach. To the best of our knowledge, this is the first time these methods have been used to calculate the correction.

\subsection{The real-space Voronoi cell method}

Our first method is based on considering the electrostatic potential directly. In such approaches, the correction is typically determined by plotting the electrostatic potential obtained from the DFT calculation, and averaging it either in the x-y plane, within atomic spheres or over primitive cells of the host material. The important point is that the averaging must be done in some localized manner, such that it is possible to measure the value of the bulk-like potential in the charged defect supercell by considering a region far away from the defect centre. The correction to the defect formation energy is then $q \left ( V_\mathrm{vac} - V_\mathrm{bulk} \right )$, where $V_\mathrm{bulk}$ and $V_\mathrm{vac}$ are the average potentials in the bulk supercell and in a `bulk-like' region of the vacancy supercell.

In order for the correction to be unambiguous and accurate, the averaging volume should be as small as possible while still covering a region of space that is completely representative of the bulk material (i.e., all regions in the bulk supercell should give the macroscopic average for the material). Our approach is to use Voronoi cells: each region corresponds to the real-space volume of points closest to one particular atom in the supercell. There are, therefore, as many regions as there are atoms. The electrostatic potential, however, is given on a discrete real-space grid whose spacing depends on the basis set cut-off energy and which, in general, is not commensurate with the atomic spacing, leading to inaccuracies in the averaging procedure that may swamp the difference between the bulk and defect supercells. In order to overcome this drawback, we Fourier-interpolate the potential onto a finer grid that is commensurate with the atomic spacing, which results in each Voronoi cell containing exactly the same number of points. We note that care must be taken to give the correct fractional weighting to points which are directly between two or more atoms.

The potential alignment correction is then determined by considering Voronoi cells belonging to the atoms which are furthest from the defect centre, as shown in Fig.~\ref{fig:pot_align}.

\subsection{The MLWF method}

Our second approach to calculating the potential alignment correction is based on matrix elements of the Hamiltonian in the basis of MLWFs. The shift in the reference of potential that is sought is reflected in the position of the energy eigenvalues; however, comparing the eigenvalues of bulk-like states in the defect supercell with those of the bulk supercell is problematic since the eigenstates are delocalized. MLWFs, on the other hand, provide a probe of the localized properties of the system. Once the Hamiltonian is transformed into a basis of MLWFs, the potential alignment correction can be determined by considering on-site matrix elements $\tilde{H}_{nn} = \langle \omega_n | \hat{H} | \omega_n \rangle$ (where $\omega_n$ is a MLWF belonging to the cell at the origin) for Wannier functions whose centres, defined by their first moments, are far away from the defect centre. The shift in potential is then simply $\tilde{H}_{nn,\mathrm{vac}} - \tilde{H}_{nn,\mathrm{bulk}}$. In practice, as shown in Fig.~\ref{fig:pot_align}, we plot this difference  for all $n$ as a function of the distance of the Wannier function centre from the defect centre, and the value of the potential alignment correction is taken as the mean of the values which are furthest from the defect centre.

This method is feasible since the wannierization procedure results in the same partitioning of the electronic density in the bulk and defect supercells everywhere except in the direct neighbourhood of the defect; hence it is possible to identify Wannier functions that are associated with the defect centre and Wannier functions that are not. In the latter case, therefore, it is straightforward to unambiguously match equivalent Wannier functions in the bulk and defect supercells. 

From Fig.~\ref{fig:pot_align} it can be seen that the MLWF correction method is in excellent agreement with the Voronoi cell method. Among all our calculations for the various charge states, the maximum discrepancy was 0.01~eV. The MLWF method gives a finer representation of the system than averaging over atomic sites (as in the Voronoi cell approach), since each silicon atom has four bonding Wannier functions connecting it to its neighbours. Therefore, there are twice as many Wannier functions as atoms, and hence twice as much information to consider.\footnote{Only the valence bands have been taken into account for the wannierization.}

MLWFs are now a standard tool for the analysis of electronic structure calculations; therefore, this approach provides a simple and accurate method for calculating the potential alignment correction, since all the necessary information is readily available once the standard wannierization procedure has been performed.\cite{wan90}

\section{Results}
\label{sec:results}

\subsection{The neutral unrelaxed vacancy}
\label{subsec:neutral_unrel}

\begin{table}
\begin{ruledtabular}
\begin{tabular*}{0.5\textwidth}{c c c c c c}
\multirow{2}{*}{$N$} & \multirow{2}{*}{Symmetry} & \multirow{2}{*}{$k_\mathrm{MP}$} & k-point vol. & \multicolumn{2}{c}{$E_f^0$ (eV)} \\
&&& ($10^{-3} \mathrm{\AA}^{-3}$) & Total & Kinetic \\
\hline
2   & FCC & 8 & 12.35 & 2.65 & -10.19\\
8   & SC  & 7 &  4.61 & 3.13 & -11.49\\
16  & FCC & 6 &  3.66 & 3.10 & -11.68\\
32  & BCC & 6 &  1.83 & 3.83 & -12.08\\
54  & FCC & 4 &  3.66 & 3.51 & -12.36\\
64  & SC  & 6 &  0.91 & 3.81 & -12.45\\
128 & FCC & 4 &  1.54 & 3.79 & -12.53\\
216 & SC  & 3 &  2.17 & 4.05 & -12.51\\
250 & FCC & 3 &  1.87 & 3.95 & -12.60\\
256 & BCC & 3 &  1.83 & 4.13 & -12.47\\
\end{tabular*}
\end{ruledtabular}
\caption{List of all supercells up to 256 atoms with their respective symmetries. Also listed are the value of $k_\mathrm{MP}$ used (converged with respect to the formation energy for each supercell), the corresponding k-point volume in reciprocal space, the unrelaxed defect formation energy and the kinetic energy contribution to this value. Calculations were performed with an ultrasoft pseudopotential and a plane-wave energy cut-off of 400~eV.\label{table:conv}}
\end{table}

In order to study the finite size convergence properties of the system we first simulate the neutral vacancy in its unrelaxed state, with all the atoms in their perfect crystalline positions; our results are reported in Table~\ref{table:conv}. Considering the different supercell geometries separately, it can be seen that $E_f^0$ increases monotonically with system size. We therefore take our highest value of 4.13~eV (for the 256-atom BCC supercell) as a lower bound to the unrelaxed defect formation energy.

The slow convergence of the formation energy as a function of system size is due to the spurious interaction between periodic images of the defect centre. As the defect is both unrelaxed and uncharged, this is neither caused by elastic interactions nor monopole--monopole electrostatic interactions; rather, it is the result of higher multipole interactions and overlap between the wavefunctions of the periodically repeated defect centres.

The importance of wavefunction overlap may be gauged by considering the kinetic energy contribution to the defect formation energy, which we define analogously to Eq.~(\ref{eqn:E_f}), except using only the non-interacting kinetic energy contribution to the total energy instead of the total energy itself. The results are listed in the right-hand column of Table~\ref{table:conv}. It is immediately apparent that the kinetic energy component of the defect formation energy varies on a scale {\em larger} than the total formation energy as the supercell size is increased, and that, even at our largest supercell sizes, it is not converged to better than 0.1~eV. These changes are caused by subtle changes in the defect wavefunctions of neighbouring defect centres that occur in order to maintain orthogonality as the periodic images of the defect move apart. While the electronic density is rather insensitive to them and, therefore, the electrostatic interactions are not affected, the kinetic energy is not. As a result, any correction scheme that accounts solely for the classical electrostatic interaction will not be sufficient in this case to predict the correct formation energy for an infinite system beyond this level of accuracy.

\begin{figure}
  \includegraphics[width=0.4\textwidth]{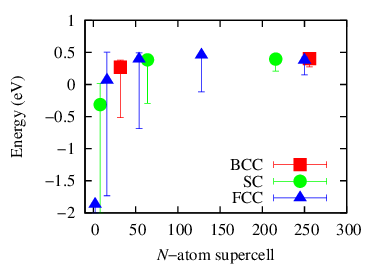}
  \caption{(Color online) Dispersion of the unrelaxed defect level. The symbols show the position of the level at $\Gamma$, and the bars show the extent of the dispersion obtained from our dense k-point sampling. The positions are given with respect to the bulk VBM.\label{fig:disp}}
\end{figure}

We consider one last question before moving on to ionic relaxations: is it desirable to use a dense k-point grid when simulating point defects? Although this is obviously necessary for describing the delocalized bulk states, it seems reasonable that a $\Gamma$-point sampling might give a better approximation of the localized defect levels in the dilute limit, as a way of deliberately eliminating unwanted dispersion, and by occupying the state which most closely resembles the bonding defect state for the infinite system\footnote{J. Neugebauer, private communication.}. Indeed, Fig.~\ref{fig:disp} shows that the position of the defect level at $\Gamma$ rapidly converges to a fixed value, even if the total dispersion of the level throughout the Brillouin zone is large.

As far as the defect formation energy is concerned, at least, our results show that $\Gamma$-point calculations do not give a better estimate of the defect formation energy: for the 256-atom supercell $E_f^0$ is underestimated (as compared to our lower bound estimate) both for the unrelaxed and relaxed cases, by 0.2~eV and 0.4~eV respectively. This conclusion is in agreement with two other studies on the relaxed silicon vacancy: Puska {\em et al.}\cite{niem-vac}, who report that a $2^3$ MP sampling shows a faster convergence than a $\Gamma$-point sampling for the neutral charge state; and Wright\cite{wright-vac}, whose results for the 216-atom supercell show a greater average error across all charge states for $\Gamma$-point sampling than any finer MP grid sampling. We note that, conversely, quantities of interest other than the defect formation energy might benefit from a $\Gamma$-point only sampling; as we shall see in Sec.~\ref{subsec:trans_levels}, the stable charge state transition levels are an example of this.

\subsection{The neutral relaxed vacancy}
\label{subsec:neutral_rel}

\begin{figure}
  \includegraphics[width=0.4\textwidth]{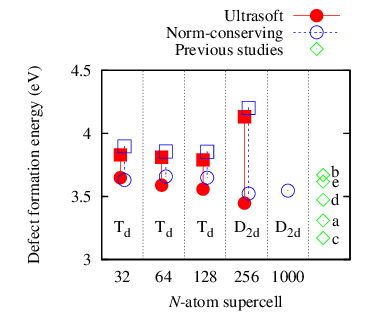}
  \caption{(Color online) Relaxation effects for the neutral vacancy. Squares show the defect formation energy for the unrelaxed lattice and circles for the relaxed one. Labelled below is the point group symmetry of the defect centre after relaxation. The last column shows the relaxed formation energy from previous studies: a) Ref.~[\onlinecite{niem-vac}] ($N=216$, $k_\mathrm{MP} = 2$, LDA); b) Ref.~[\onlinecite{niem-multisymm}] ($N=216$, $k_\mathrm{MP} = 2$, GGA); c) Ref.~[\onlinecite{probert}] ($N=256$, $k_\mathrm{MP} = 2$, GGA); d) Ref.~[\onlinecite{wright-vac}] ($N=1000$, $k_\mathrm{MP} = 3$, LDA); e) Ref.~[\onlinecite{wright-vac}] ($N=1000$, $k_\mathrm{MP} = 3$, GGA). Our calculations were performed with an ultrasoft (norm-conserving) pseudopotential and the LDA functional, with a converged plane-wave energy cut-off of 400~eV (800~eV) and a k-point grid of $k_\mathrm{MP} = 3$ ($k_\mathrm{MP} = 4$).\label{fig:rel}}
\end{figure}

On relaxation of the ionic positions, the predicted symmetry for $\mathrm{v}^0$ from Watkins' model ($\mathrm{D_{2d}}$) is not seen in any supercell smaller than 256 atoms, as shown in Fig.~\ref{fig:rel}. Calculations on the 32-atom BCC, 64-atom SC and 128-atom FCC supercells do not exhibit a change in symmetry, only an inwards relaxation. Consequently, the defect formation energy is only lowered by $\sim$0.2~eV in these cases.

The 256-atom BCC supercell undergoes the predicted change in symmetry and a reduction in defect volume by more than $40\%$. This results in the defect formation energy being lowered by 0.7~eV (full results are given in Table~\ref{table:summary}). Therefore, even though the unrelaxed defect formation energy is highest for the largest supercell, after relaxation it becomes the lowest, demonstrating the well-known importance of Jahn-Teller distortion and the long-ranged nature of the elastic interactions which can lead to both qualitatively incorrect relaxation patterns and quantitatively inaccurate formation energies in small supercells.

These results confirm several previous DFT studies on this system\cite{probert,niem-vac,wright-vac}. In particular, Puska {\em et al.}\cite{niem-vac} report that small supercells can show a range of different symmetries depending on the k-point sampling, and there is some evidence from their results that a $\Gamma$-point only sampling favours $\mathrm{D_{2d}}$. It is clear from our results that a dense k-point sampling favours the lattice's unrelaxed tetrahedral point group symmetry $\mathrm{T_d}$ for small supercells.

Fig.~\ref{fig:rel} also shows the relaxed defect formation energy in a 1000-atom SC supercell, as calculated with the linear-scaling DFT code {\sc onetep}.\footnote{The unrelaxed defect formation energy has not been calculated with {\sc onetep}, since the density matrix method does not allow for a degenerate ground state without an explicit (non-linear-scaling) diagonalization.} The result (calculated with the same pseudopotential) is within 23~meV of the 256-atom BCC system, and the relaxed defect structure is almost identical, with only a 33~$\mathrm{m\AA}$ root mean square difference in the lengths of the bonds at the defect centre.

\subsection{Charged vacancies}
\label{subsec:charged}

\begin{table*}
\begin{ruledtabular}
\begin{tabular*}{0.5\textwidth}{c c c c c c c c c c c c}
\multirow{2}{*}{Charge} & \multirow{2}{*}{$N$} & \multirow{2}{*}{$k_\mathrm{MP}$} & \multirow{2}{*}{Relaxed} & \multirow{2}{*}{Pseudopotential} & \multirow{2}{*}{Symmetry} & \multicolumn{3}{c}{Bond lengths ($\mathrm{\AA}$)} & \multirow{2}{*}{Defect vol. ($\mathrm{\AA}^3$)} & \multicolumn{2}{c}{$E_f^q$ (eV)} \\
&&&&&& a--a & b--b & a--b && Hole & VBM \\
\hline
\multirow{2}{*}{$2+$} & 256 & 1 & Yes & Ultrasoft & $\mathrm{T_d}$ & 3.22 & 3.22 & 3.22 & 3.93 & 2.86 & 2.85 \\
                      & 256 & 3 & Yes & Ultrasoft & $\mathrm{T_d}$ & 3.63 & 3.63 & 3.63 & 5.61 & 3.62 & 3.66 \\
\hline
     & 256 & 1 & Yes & Ultrasoft & $\mathrm{D_{2d}}$ & 2.93 & 2.93 & 3.27 & 3.62 & 2.99 & 3.00 \\
$1+$ & 256 & 3 & Yes & Ultrasoft & $\mathrm{D_{2d}}$ & 2.86 & 2.86 & 3.40 & 3.73 & 3.50 & 3.55 \\
     & 1000 & 1 & No\footnote{For this system we use the relaxed atomic positions obtained for the 1000-atom $\mathrm{v}^0$ supercell, without further relaxation.} & Norm-conserving & $\mathrm{D_{2d}}$ & 2.87 & 2.87 & 3.37 & 3.70 & - & 3.50 \\
\hline
    & 256 & 1 & No & Ultrasoft & $\mathrm{T_d}$ & 3.81 & 3.81 & 3.81 & 6.54 & \multicolumn{2}{c}{3.90} \\
    & 256 & 3 & No & Ultrasoft & $\mathrm{T_d}$ & 3.81 & 3.81 & 3.81 & 6.54 & \multicolumn{2}{c}{4.13} \\
    & 256 & 1 & Yes & Ultrasoft & $\mathrm{D_{2d}}$ & 2.83 & 2.83 & 3.31 & 3.53 & \multicolumn{2}{c}{3.06} \\
$0$ & 256 & 3 & Yes & Ultrasoft & $\mathrm{D_{2d}}$ & 2.86 & 2.86 & 3.40 & 3.73 & \multicolumn{2}{c}{3.45} \\
    & 256 & 4 & No & Norm-conserving & $\mathrm{T_d}$ & 3.81 & 3.81 & 3.81 & 6.50 & \multicolumn{2}{c}{4.20} \\
    & 256 & 4 & Yes & Norm-conserving & $\mathrm{D_{2d}}$ & 2.86 & 2.86 & 3.41 & 3.74 & \multicolumn{2}{c}{3.52} \\
    & 1000 & 1 & Yes & Norm-conserving & $\mathrm{D_{2d}}$ & 2.87 & 2.87 & 3.37 & 3.70 & \multicolumn{2}{c}{3.55} \\
\hline
\multirow{2}{*}{$1-$} & 256 & 1 & Yes & Ultrasoft & $\sim$$\mathrm{C_{2v}}$ & 3.10 & 2.70 & 3.24 (av.) & 3.50 & 3.48 & 3.44 \\
                      & 256 & 3 & Yes & Ultrasoft & $\sim$$\mathrm{C_{2v}}$ & 2.94 & 2.87 & 3.26 (av.) & 3.55 & 3.80 & 3.71 \\
\hline
&&&&&& a--a & a--c & $\Delta \alpha$ && \\
\multirow{2}{*}{$2-$} & 256 & 1 & Yes & Ultrasoft & $\mathrm{D_{3d}}$ (split) & 3.45 & 2.60 & $6.63^{\circ}$ & 2.86 & 3.70 & 3.57  \\
                      & 256 & 3 & Yes & Ultrasoft & $\mathrm{D_{3d}}$ (split) & 3.46 & 2.60 & $6.52^{\circ}$ & 2.88 & 3.92 & 3.72 \\
\end{tabular*}
\end{ruledtabular}
\caption{Summary table of the results obtained for the 256-atom BCC supercell and the 1000-atom SC supercell. Bond lengths are given between the atoms surrounding the vacancy, as labelled in Fig.~\ref{subfig:label-watkins} for $\mathrm{v}^{2+}$, $\mathrm{v}^{1+}$, $\mathrm{v}^0$, $\mathrm{v}^{1-}$, and Fig.~\ref{subfig:label-split} for $\mathrm{v}^{2-}$. Atoms labelled with the same letter are equivalent by symmetry. $\Delta \alpha = \alpha - 90^{\circ}$ is the distortion of the regular octahedron in the split vacancy configuration. The defect volume is calculated from the tetrahedron formed by the four neighbours of the vacancy site. The defect formation energy is given for $\Delta \mu_e = 0$. Symmetry labels are given in Sch\"{o}nflies notation.\label{table:summary}}
\end{table*}

Our results for the relaxed symmetries and defect formation energies of the various charge states of the vacancy are shown in Table~\ref{table:summary}. All the charge states apart from $\mathrm{v}^{2-}$ are in agreement with Watkins' model. For $\mathrm{v}^{2-}$ a completely different `split vacancy' configuration is found, with a point group symmetry of $\mathrm{D_{3d}}$ (in agreement with previous studies by Nieminen {\em et al.}\cite{niem-vac,niem-multisymm} and Wright\cite{wright-vac}). The symmetry predicted by the model ($\mathrm{C_{2v}}$) was found as a metastable state. For all charge states there were no qualitative differences between the $\Gamma$-point only and multiple k-point calculations, although the former consistently showed a greater inwards relaxation and thus a smaller defect volume. For a detailed discussion of the electronic structure of the different charge states of the vacancy, the reader is referred to Appendix~\ref{app:res-mlwfs}.

We note that the relaxation of $\mathrm{v}^{1-}$ is the most problematic; in fact, if the initial random displacement of the atomic positions is not large enough, the system consistently converges to a metastable $\mathrm{T_d}$ configuration with a completely spin-polarized triply degenerate defect level (i.e., filled with three spin-aligned electrons). Furthermore, our final lowest configuration is only approximately $\mathrm{C_{2v}}$, as there is a small but noticeable splitting of $\sim$0.1~$\mathrm{\AA}$ of the bond lengths between the two pairs of neighbours of the vacancy. Nieminen and Wright both report that the LDA does not give the expected symmetry for this charge state, although they find a $\mathrm{D_{3d}}$ configuration.

As explained in Sec.~\ref{sec:pot_align}, there are two possible approaches for estimating the position of $\varepsilon_v$ in the system: we can either use a calculation of the bulk supercell with an electron hole (denoted `Hole' in Tables~\ref{table:summary}--\ref{table:levels}), or the bulk VBM value with a potential alignment correction\footnote{The results given use the Voronoi cell method for determining the potential alignment; this is to ensure consistency among all results, since for the 1000-atom {\sc onetep} calculation only the Voronoi cell method is appropriate. For all comparisons of {\sc castep} calculations between the Voronoi cell and MLWFs methods the agreement was to within 0.01~eV (calculations not shown).} (`VBM'). The agreement between the two approaches is better for the positively charged vacancies ($-0.05$~eV to 0.01~eV) than the negatively charged ones (0.05~eV to 0.21~eV); however, these uncertainties are small enough not to affect the level ordering.

\subsection{Transition levels}
\label{subsec:trans_levels}

\begin{table*}
\begin{ruledtabular}
\begin{tabular*}{0.5\textwidth}{l c c c c c c c c c c}
& \multirow{4}{*}{Expt.\cite{watkins-model}} &&&&&& Ref.~[\onlinecite{niem-vac}] & Ref.~[\onlinecite{niem-multisymm}]\footnote{The transition levels have been calculated from the quoted values of the formation energy of the various charge states using Eq.~(\ref{eqn:levels}).} & \multicolumn{2}{c}{Ref.~[\onlinecite{wright-vac}]$^\mathrm{a}$} \\
&& \multicolumn{4}{c}{$N = 256$} & $N = 1000$ & $N = 216$ & $N = 216$ & \multicolumn{2}{c}{$N = 1000$} \\
&& \multicolumn{2}{c}{$k_\mathrm{MP} = 1$} & \multicolumn{2}{c}{$k_\mathrm{MP} = 3$} & $k_\mathrm{MP} = 1$ & $k_\mathrm{MP} = 1$ & $k_\mathrm{MP} = 2$ & \multicolumn{2}{c}{$k_\mathrm{MP} = 3$} \\
&& Hole & VBM & Hole & VBM & VBM & LDA & GGA & LDA & GGA \\
\hline
$E \left ( {2+} / {1+} \right )$ & 0.13  & 0.13  & 0.16  & $\times$ & $\times$ & -    & 0.19  & 0.13  & 0.27* & 0.27  \\
$E \left ( {2+} / 0 \right )$    & 0.09* & 0.10* & 0.11* & $\times$ & $\times$ & -    & 0.15* & 0.10* & 0.28  & 0.19* \\
$E \left ( {1+} / 0 \right )$    & 0.05  & 0.07  & 0.06  & $\times$ & $\times$ & 0.04 & 0.11  & 0.06  & 0.28* & 0.12  \\
$E \left ( 0 / {1-} \right )$    & -     & 0.43  & 0.38  & 0.35     & 0.26     & -    & 0.57  & 0.37  & 0.76  & 0.63  \\
$E \left ( 0 / {2-} \right )$    & -     & 0.32* & 0.25* & 0.24*    & 0.13*    & -    & 0.49* & 0.37* & 0.66* & 0.53* \\
$E \left ( {1-} / {2-} \right )$ & -     & 0.22  & 0.13  & 0.13     & 0.00     & -    & 0.40  & 0.36  & 0.56  & 0.42  \\
\end{tabular*}
\end{ruledtabular}
\caption{Transition levels for the 256-atom BCC supercell and the 1000-atom SC supercell. Results from previous studies are also shown. Our calculations use the LDA functional. Crosses ($\times$) are used when there is no transition in the band gap. Asterisks (*) denote thermodynamically stable transitions. All values are given in eV.\label{table:levels}}
\end{table*}

\begin{figure}
  \includegraphics[width=0.4\textwidth]{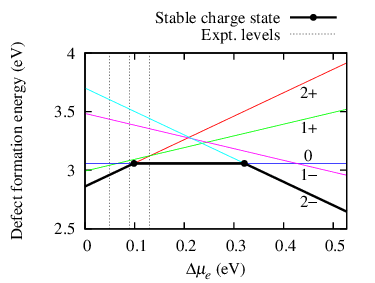}
  \caption{(Color online) Formation energy of the different charge states of the vacancy as a function of the electron chemical potential (plotted relative to the VBM). The energy range shown covers the DFT band gap for silicon. The thermodynamically stable charge state at each point is highlighted in bold, and the circles indicate the level position for the stable transitions. The dashed vertical lines show the available experimental values for the transition levels. The results shown are for the 256-atom BCC supercell with a $\Gamma$-point sampling.\label{fig:stable}}
\end{figure}

Following the careful description of Baraff {\em et al.}\cite{negU-baraff}, we define the stable charge state transition level $E \left ( m / n \right )$ to be the value of the Fermi level (with respect to the perfect crystal valence band edge) at which there is a crossing of the defect formation energies of two charge states $m$ and $n$, leading to a change in the most stable state from $\mathrm{v}^m$ to $\mathrm{v}^n$ as the Fermi level is raised. This is given by
\begin{equation}\label{eqn:levels}
E \left ( m / n \right ) = \frac{E_f^n \left ( \Delta \mu_e = 0 \right ) - E_f^m \left ( \Delta \mu_e = 0 \right ) }{m - n}.
\end{equation}

As first predicted by Baraff {\em et al.}\cite{negU-baraff} and then experimentally verified by Watkins and Troxell\cite{negU}, the isolated silicon vacancy exhibits a `negative-U' effect, by which the stable charge state changes directly from $\mathrm{v}^{2+}$ to $\mathrm{v}^0$. The $\mathrm{v}^{1+}$ state is therefore only metastable; this is a direct consequence of the Jahn-Teller distortion, which lowers the energy of the neutral vacancy much more than the singly positive vacancy. It has also been suggested that the same effect might be observed in the sequence $\mathrm{v}^0$, $\mathrm{v}^{1-}$, $\mathrm{v}^{2-}$.

Our transition levels are given in Table~\ref{table:levels}; as can be seen, the $\Gamma$-point calculations predict the negative-U behaviour both for the negatively and positively charged levels as expected, and are in good agreement with the available experimental results (also shown in Fig.~\ref{fig:stable}). Surprisingly, the calculations using a dense k-point sampling ($k_\mathrm{MP} = 3$) give the opposite ordering for the sequence $\mathrm{v}^{2+}$, $\mathrm{v}^{1+}$, $\mathrm{v}^0$; this results in no transitions in the band gap between these levels, and only $\mathrm{v}^0$ being a stable charge state. The negative-U behaviour is however still present for the negatively charged levels, although their positioning is lower than for the $\Gamma$-point calculations. The $\Gamma$-point, therefore, gives better estimates of the transition levels. However, as stated previously, the absolute value of the defect formation energy is not as well converged with respect to the dilute limit when using this Brillouin zone sampling. However, calculations of $\mathrm{v}^0$ and $\mathrm{v}^{1+}$ in the 1000-atom supercell at $\Gamma$ (also included in Tables~\ref{table:summary}--\ref{table:levels}) are converged with respect to system size both in terms of the transition level $E \left ( {1+} / 0 \right ) = 0.04~\mathrm{eV}$ and the absolute value of the formation energy $E_f^0 = 3.55~\mathrm{eV}$. Therefore, this suggests that at this system size a $\Gamma$-point sampling can be employed for simultaneous convergence of the both quantities of interest.

\section{Conclusions}
\label{sec:conclusions}

We have studied the silicon vacancy in all its charge states using the supercell approach and plane-wave pseudopotential DFT. Our calculations confirm the slow finite size convergence of defect formation energies and transition levels, due to electrostatic interactions and wavefunction overlap between periodic images of the defect, and long-ranged ionic relaxations. The impact of each of these has been quantified, and it has been found that all three provide non-negligible contributions to the total error. Furthermore, due to the hybridization of the defect levels with the perfect crystal band structure, we find that the choice of k-points has a noticeable impact on the results. In particular, $\Gamma$-point calculations converge faster than calculations with uniform, multiple k-point sampling when considering stable charge state transition levels (given by differences in formation energies of different charge states), and vice-versa when considering the absolute value of formation energies.

Due to this slow convergence, the calculations may still benefit from the use of even larger supercells than the ones currently employed. This, however, will present additional numerical challenges: the defect formation energy is an {\em intensive} quantity that is obtained by taking the difference of two total energies that are {\em extensive} with the system size. As a result, to achieve a given accuracy for $E_f$ as the supercell size is increased, it is necessary to proportionately increase the precision per atom in the total energy calculations in order to avoid the result being swamped by numerical noise.

We have also introduced two methods for correcting the alignment of the valence band maximum for charged defects: one is based on averaging the electrostatic potential using a Voronoi cell construction, and the other on Hamiltonian matrix elements in a basis of maximally-localized Wannier functions. The two methods give excellent mutual agreement and constitute simple and robust ways to calculate potential alignment corrections. However, the determination of $\varepsilon_v$ remains a somewhat uncontrolled source of error, as demonstrated by the discrepancy between the potential alignment correction (`VBM') and total energy difference (`Hole') approaches (described in Sec.~\ref{sec:pot_align}), which can be as small as 0.01~eV (for positively charged defects), or as large as 0.2~eV (for negatively charged defects), as shown in Table~\ref{table:summary} (right hand columns). We note that the problem of correcting errors in charged supercells arising from periodic boundary conditions has been addressed by many previous studies\cite{interactions,gap-prob,def_corr,hybrid2,mp-over2,nick-supercell_shape}.

The accuracy of the LDA functional for describing exchange and correlation should also be investigated. GGA calculations have already been shown to give very similar qualitative and quantitative results (as shown in Sec.~\ref{sec:results}); both local and semi-local functionals, however, suffer in particular with respect to the well-known effect of self-interaction error. Unfortunately, higher accuracy methods are at present confined to relatively small system sizes due their computational expense, and are therefore limited by the large finite size errors that we have discussed earlier. Nonetheless, qualitative differences have been suggested for the vacancy from screened-exchange calculations.\cite{sx_lda-vac}

Finally, it is interesting to note that maximally-localized Wannier functions give a chemically intuitive picture of the electronic structure at the vacancy site, and without any external input confirm Watkins' LCAO model of the defect. The only charge state which does not follow the predictions from the model (the doubly negative vacancy) is also explained in terms of its Wannier functions. Full details are given in Appendix~\ref{app:res-mlwfs}.

\begin{acknowledgments}
This work was supported by the UK Engineering and Physical Sciences Research Council (EPSRC). The calculations were performed on cx1/cx2 (Imperial College London High Performance Computing Service) and HECToR (UK National Supercomputing Service). We thank M.~J.~Puska and R.~M.~Nieminen for helpful discussions, and the UK's HPC Materials Chemistry Consortium for access to HECToR.
\end{acknowledgments}

\appendix

\section{Visualization using MLWFs}
\label{app:res-mlwfs}

Wannier functions provide a localized view of the electronic structure. This is particularly useful in the study of point defects, since the Bloch states associated with the defect levels are not completely localized due to the periodicity of the system; in fact, in small supercells it is not possible to disentangle them from the bulk band structure.

\begin{figure*}
  \subfigure[\ Tetrahedral lattice schematic\label{subfig:label-watkins}]{\includegraphics[width=0.3\textwidth]{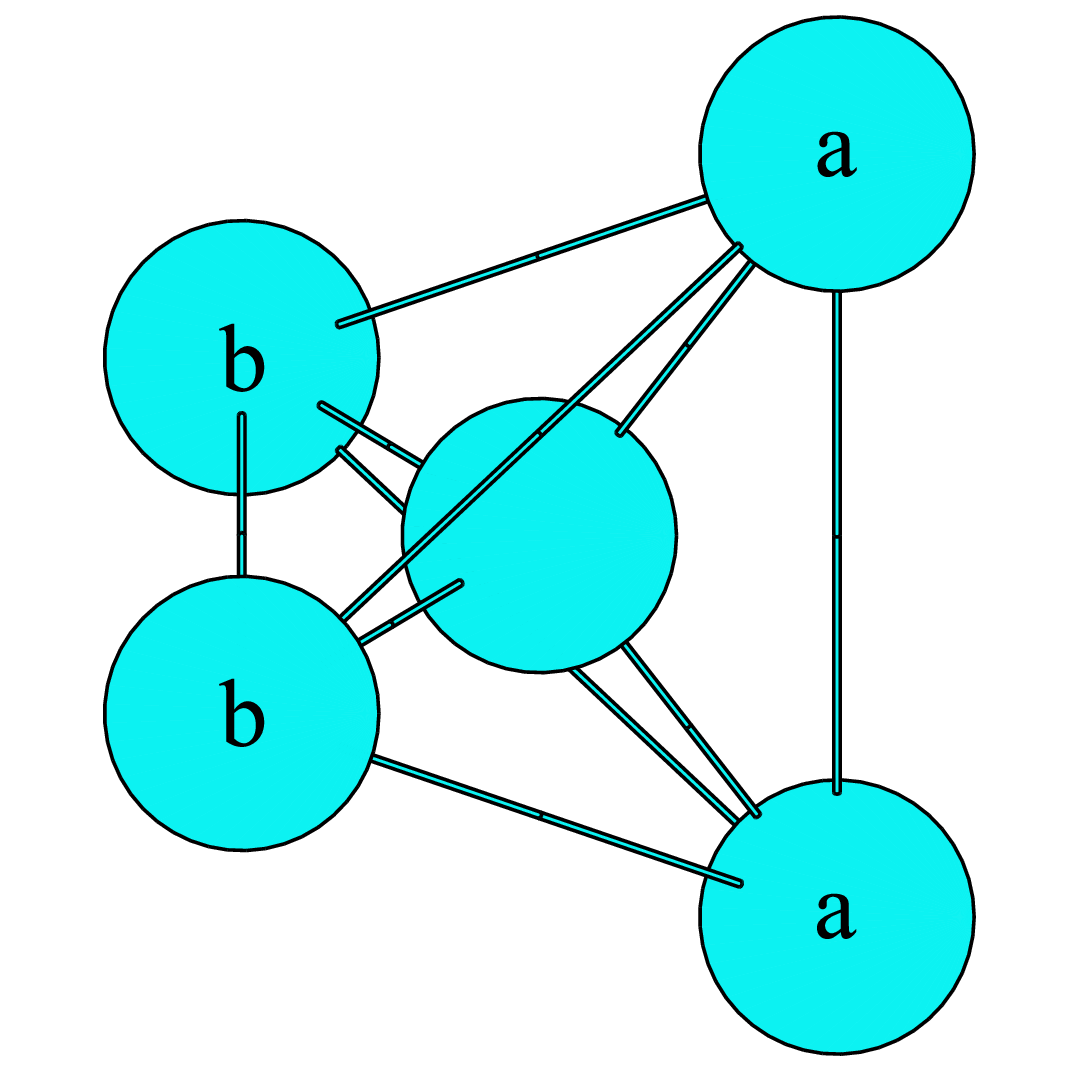}}
  \subfigure[\ Bulk Si; $A = 2.5/\sqrt{v}$, $f = 2$\label{subfig:wan-bulk}]{\includegraphics[width=0.3\textwidth]{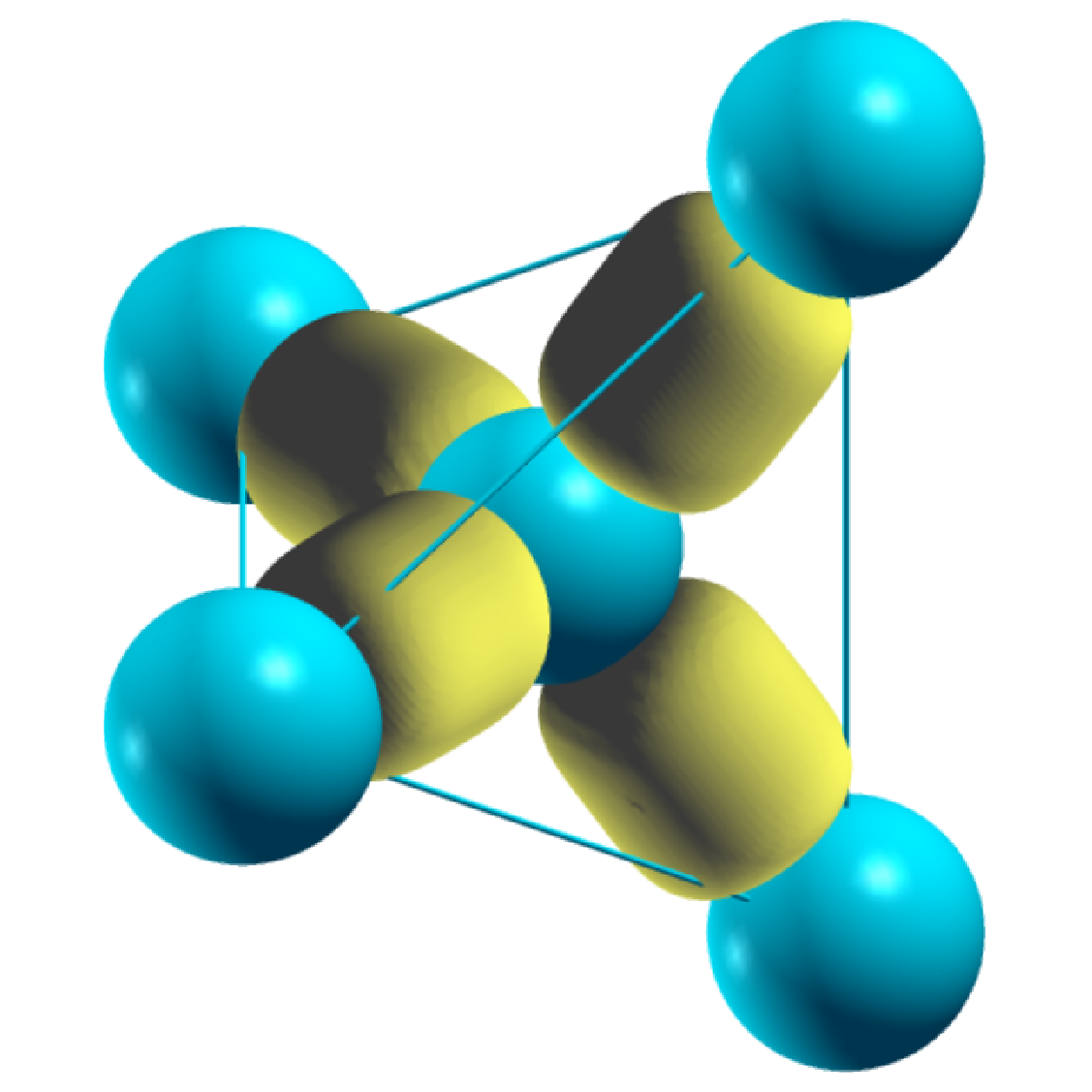}}
  \subfigure[\ $\mathrm{v}^0$ (unrelaxed); $A = 2/\sqrt{v}$, $f \approx 1$\label{subfig:wan-unrel}]{\includegraphics[width=0.3\textwidth]{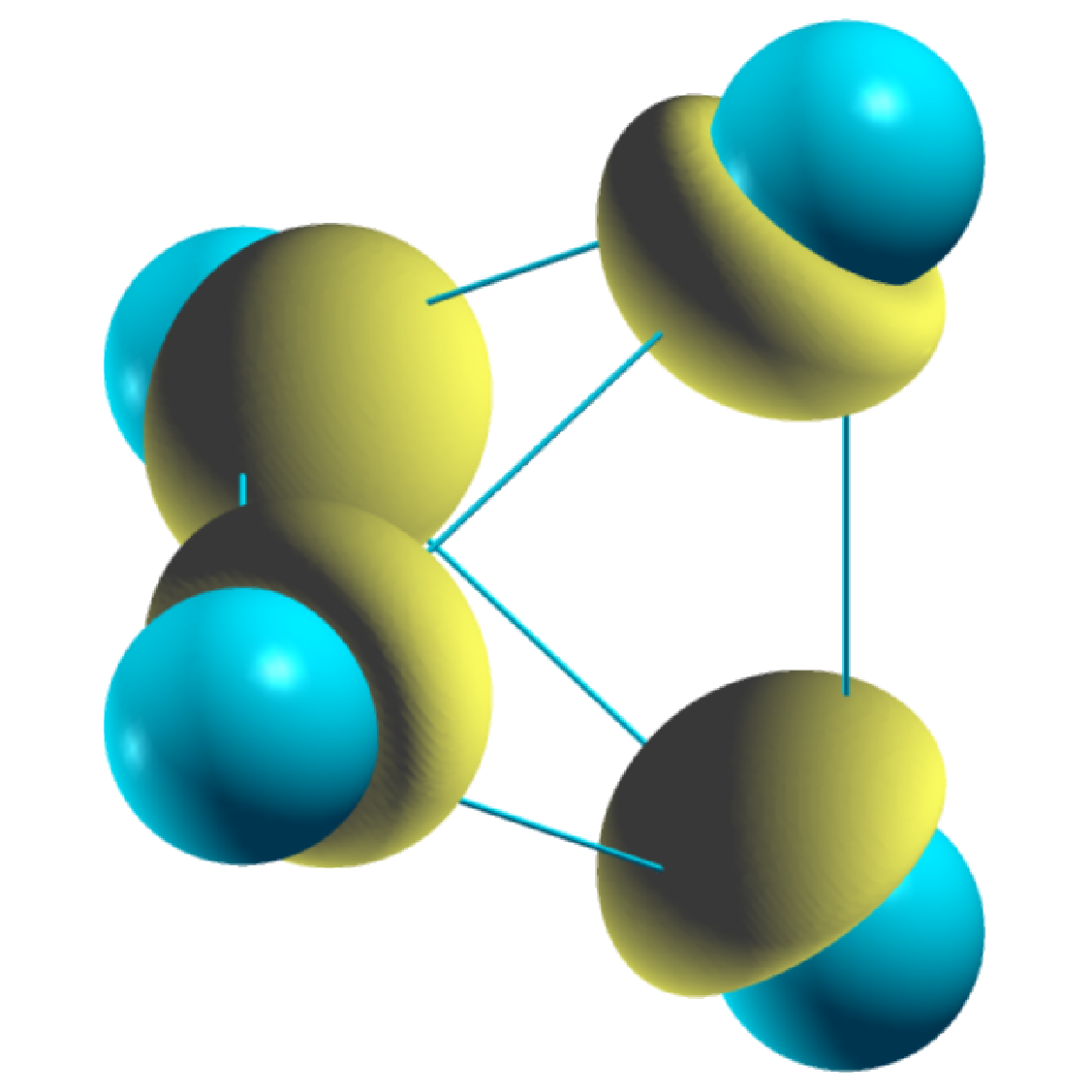}}
  \subfigure[\ $\mathrm{v}^{2+}$ (relaxed); $A = 0.75/\sqrt{v}$, $f = 2$\label{subfig:wan-2+}]{\includegraphics[width=0.3\textwidth]{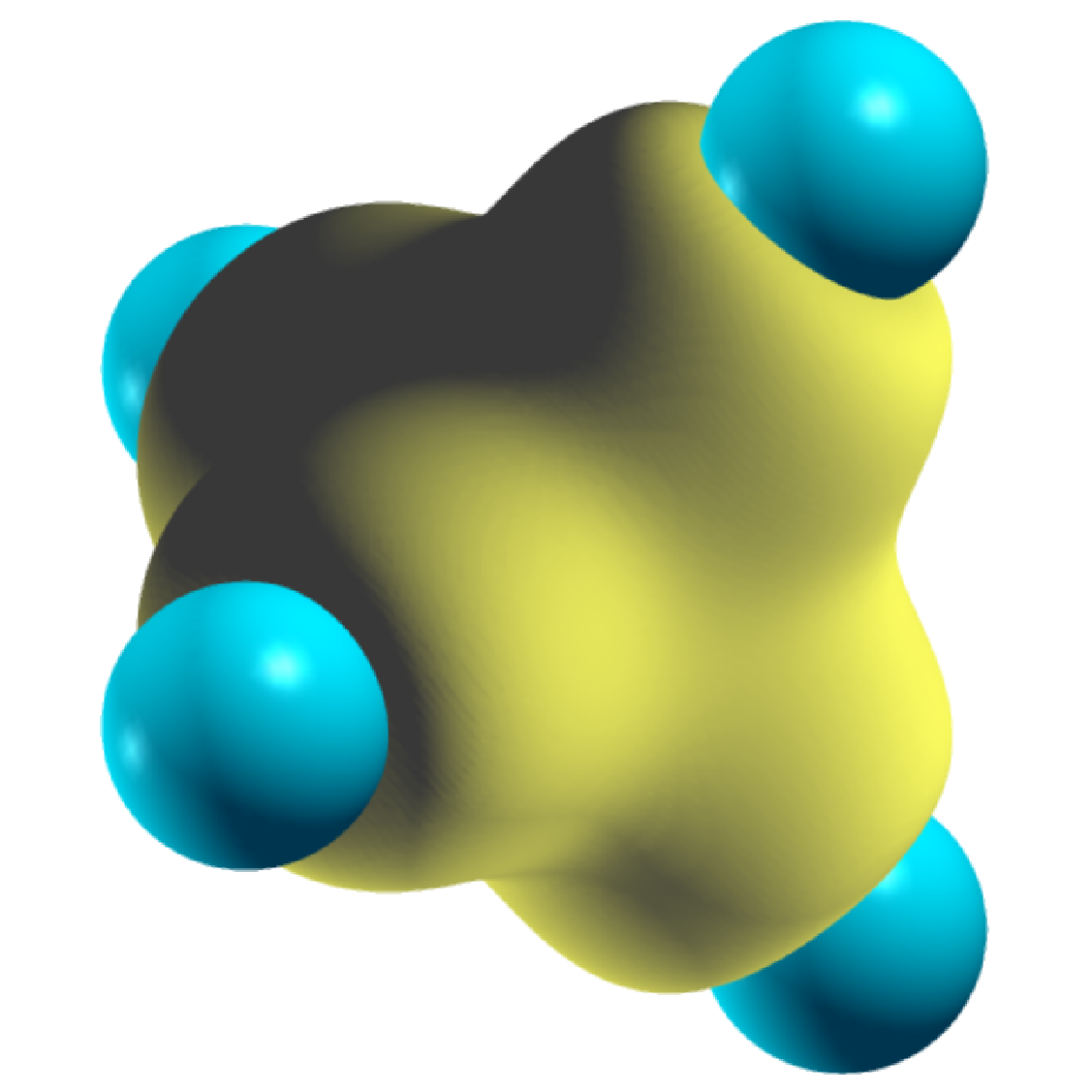}}
  \subfigure[\ $\mathrm{v}^0$ (relaxed); $A = 1.5/\sqrt{v}$, $f = 2$\label{subfig:wan-0}]{\includegraphics[width=0.3\textwidth]{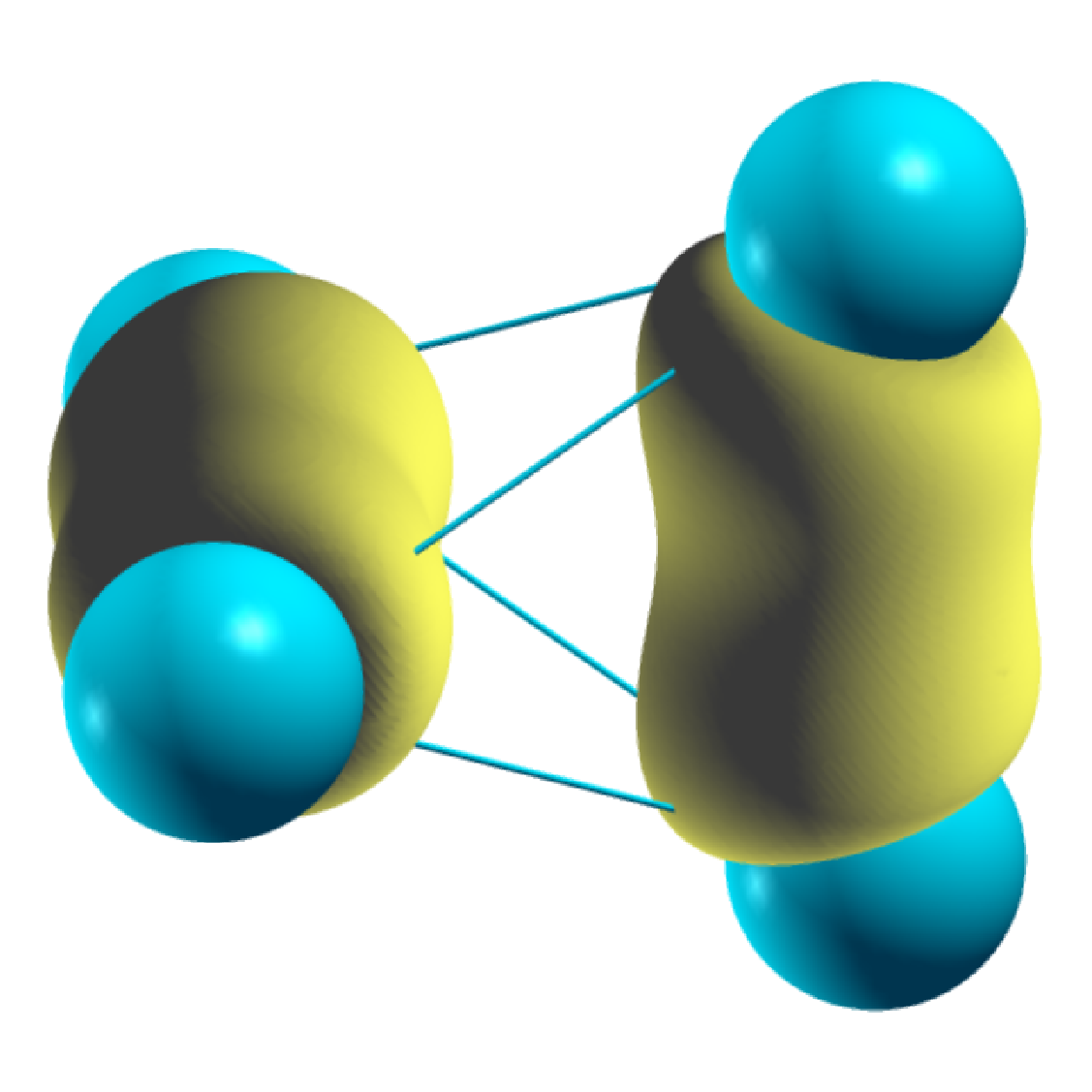}}
  \subfigure[\ $\mathrm{v}^{1-}$ (relaxed); $A = 1.75/\sqrt{v}$, $f = 2$\label{subfig:wan-1-}]{\includegraphics[width=0.3\textwidth]{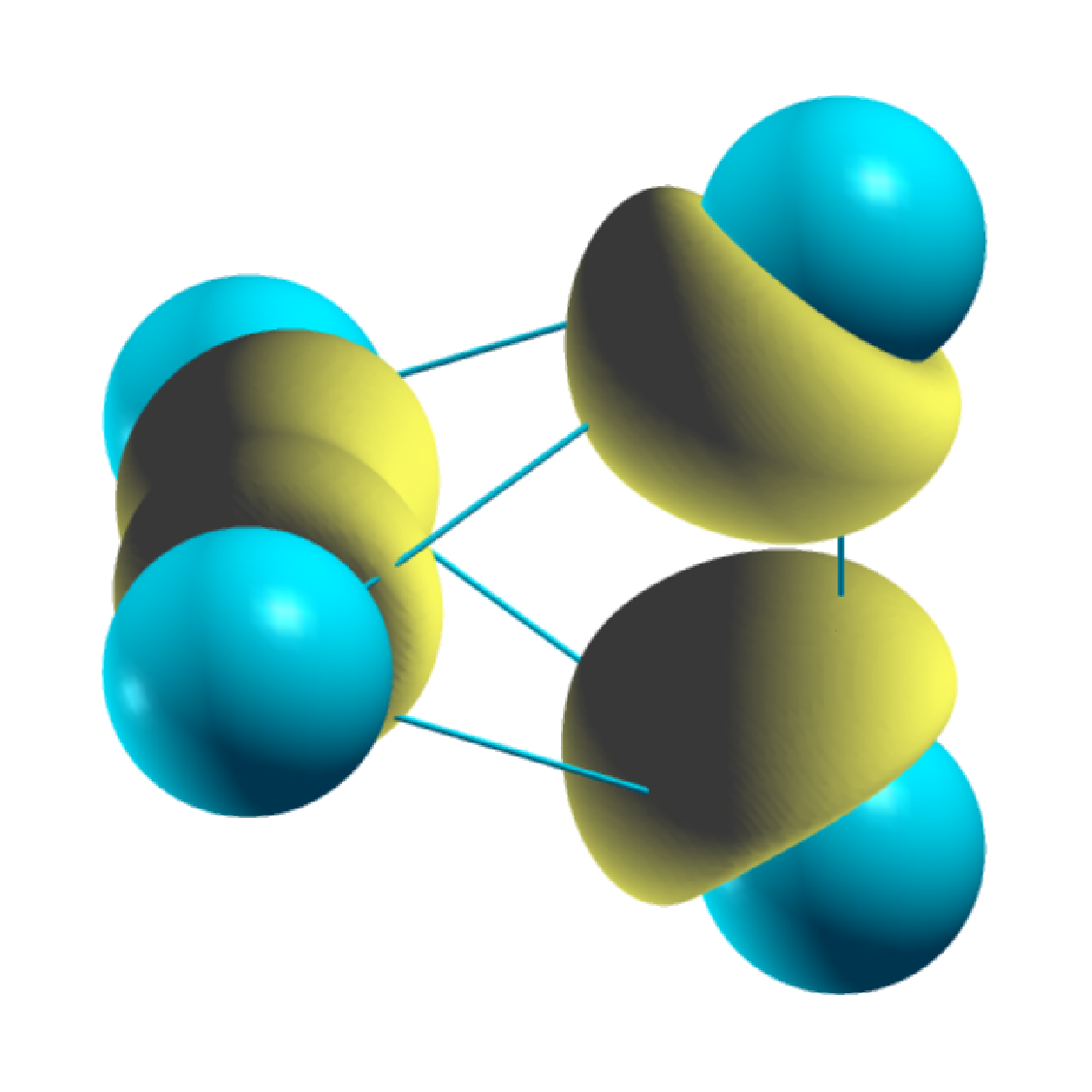}}
  \subfigure[\ Split vacancy lattice schematic\label{subfig:label-split}]{\includegraphics[width=0.3\textwidth]{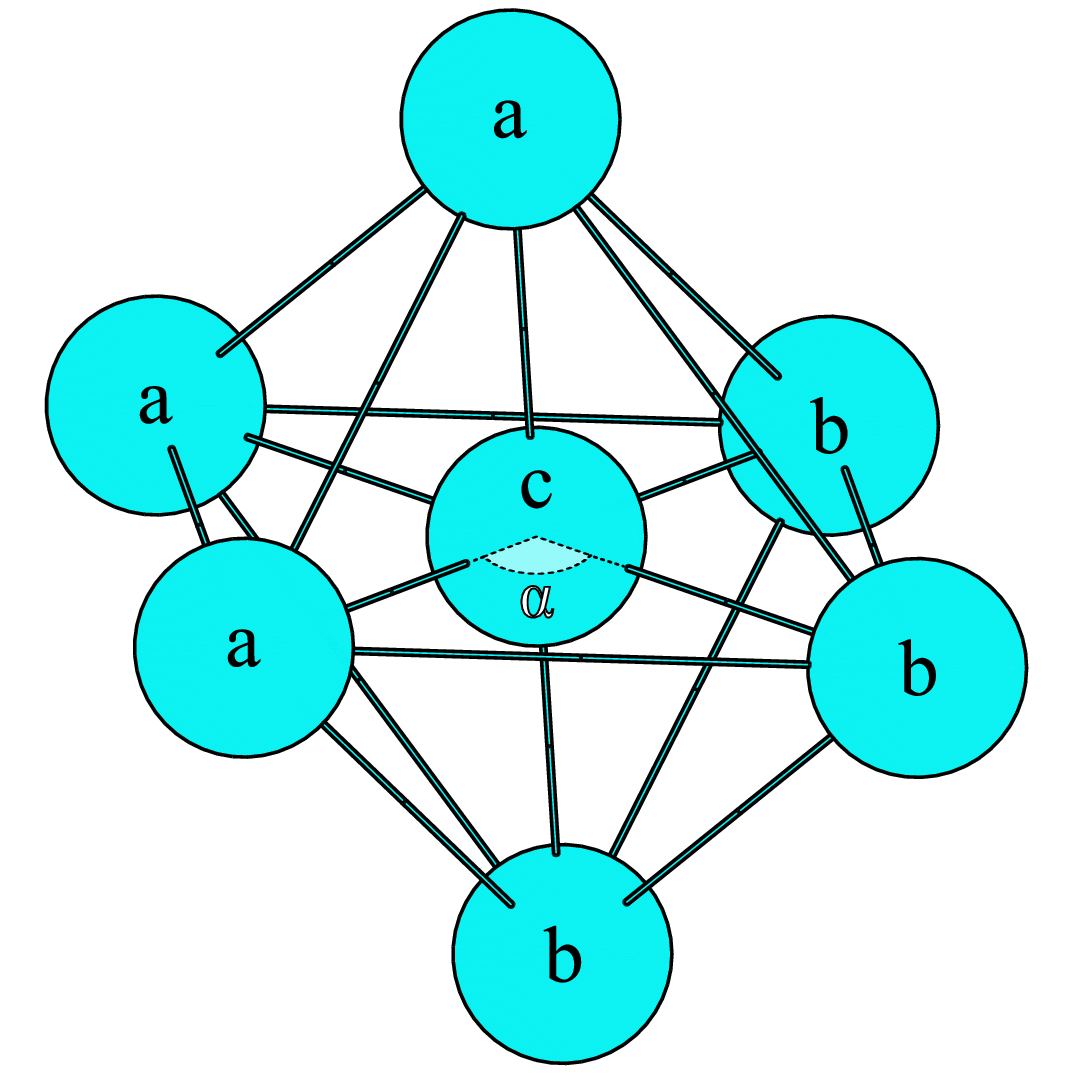}}
  \subfigure[\ $\mathrm{v}^{2-}$ (relaxed); $A = 2.25/\sqrt{v}$, $f = 2$\label{subfig:wan-2-}]{\includegraphics[width=0.3\textwidth]{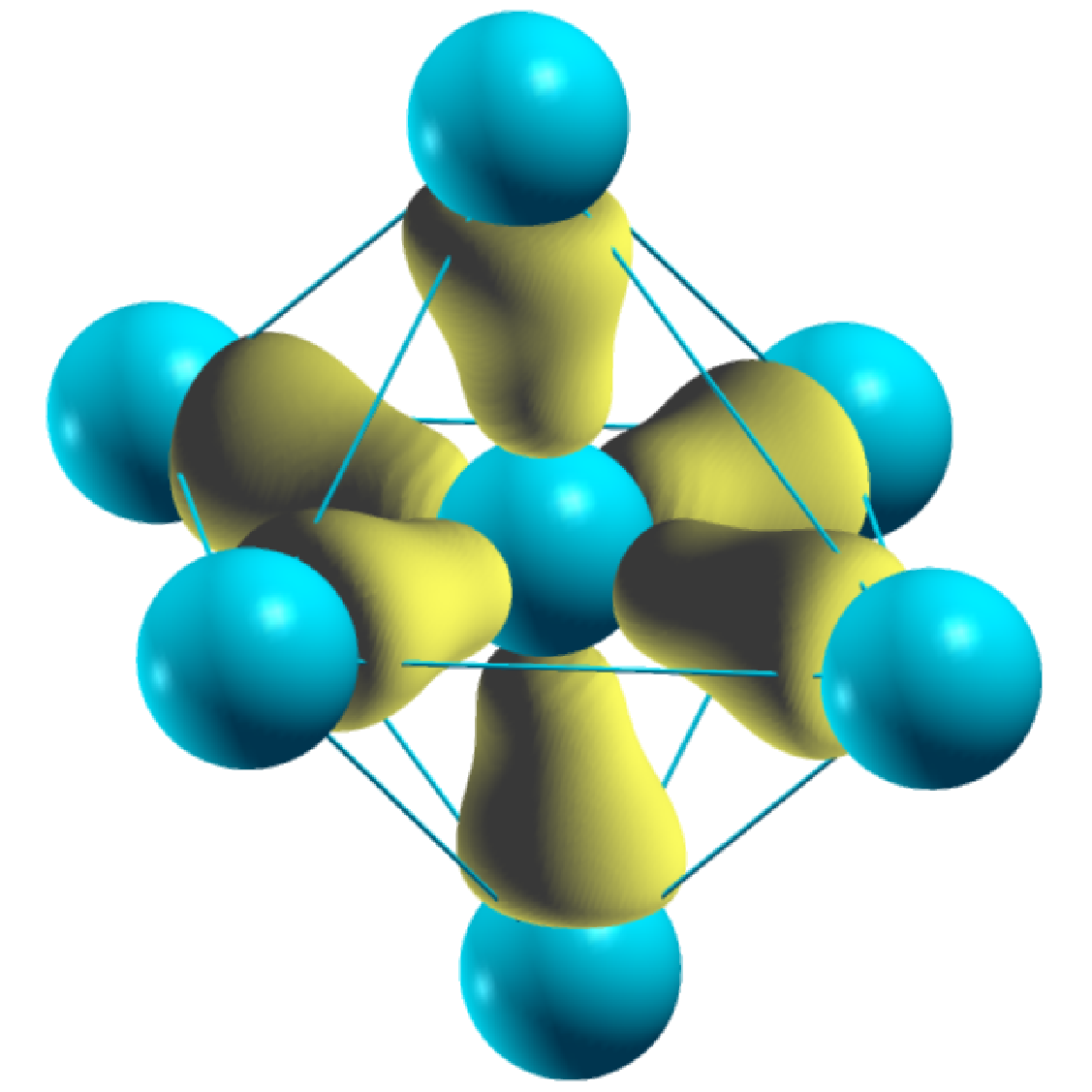}}
  \caption{(Color online) Contour-surface plots of the MLWFs most strongly associated with the defect centre for different charge states of the vacancy, calculated using the 256-atom BCC supercell and a norm-conserving pseudopotential; for computational efficiency we only include the $\Gamma$-point wavefunction in the wannierization procedure, since we expect a negligible difference in the qualitative features of the resulting Wannier functions. Each separate closed surface is an individual Wannier function. The amplitude $A$ of the contour shown in each figure is given in the caption in terms of the primitive cell volume $v$. In general the Wannier functions also have small negative components on the backbonds (not pictured). $f$ is the electronic occupancy of a single orbital. The wannierization procedure is performed using the {\sc wannier90}\cite{wan90} code (version 1.2).\label{fig:wan}}
\end{figure*}

In the case of bulk silicon, previous work has shown that it is possible to recover the typical $\sigma$ bond orbitals between silicon ions by wannierization of the valence band; alternatively, the bottom conduction bands (once disentangled from higher bands) give the corresponding antibonding orbitals, while treating both these sets of bands together as a single manifold produces four $sp^3$ orbitals on each ion.\cite{mlwf,wan-ent} The bulk silicon $\sigma$ orbitals obtained from our calculations are shown in Fig.~\ref{subfig:wan-bulk}.

For the wannierization of the defect systems we only use the occupied manifold, except in the case of the unrelaxed vacancy where we include all the defect levels. We can loosely define the concept of a defect Wannier function to be any MLWF in the defect supercell which differs qualitatively from the $\sigma$ bond MLWFs of the bulk system. For all charge states, such defect Wannier functions are only present within the first ionic shell around the vacancy; between the first and second shell we already recover $\sigma$-like bonding orbitals.

\begin{figure}
  \includegraphics[width=0.45\textwidth]{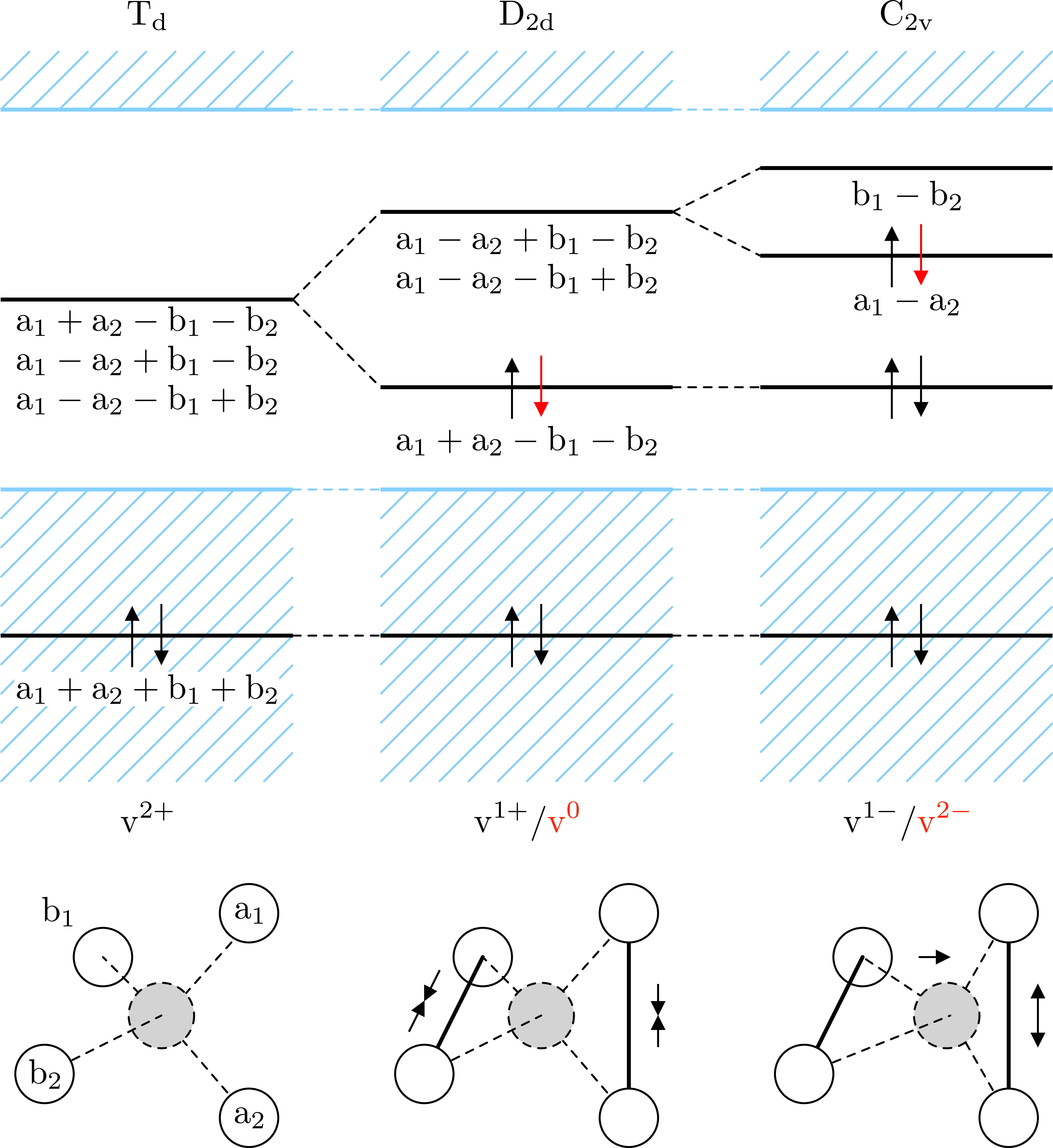}
  \caption{(Color online) Watkins' LCAO model for the silicon vacancy, deduced from electron paramagnetic resonance (EPR) studies of the defect centre\cite{jt-exp}, and later confirmed by electron-nuclear double resonance (ENDOR) measurements\cite{endor1,endor2}. The four orbitals associated with the neighbouring atoms of the defect site are the dangling bonds resulting from the removal of the central silicon atom. The figure shows the predicted ionic configuration and point group symmetry for different charge states of the vacancy due to Jahn-Teller distortion\cite{jt-theorem}. For the $\mathrm{D_{2d}}$ and $\mathrm{C_{2v}}$ configurations, the first charge state (in black) refers to the spin-polarized case with only one electron in the highest occupied orbital, and the second charge state (in red) refers to the non-spin-polarized case with two electrons.\label{fig:split}}
\end{figure}

Fig.~\ref{subfig:wan-unrel} shows the defect Wannier functions for the neutral unrelaxed system; as expected from Watkins' model (described in Fig.~\ref{fig:split}), these are $sp^3$ orbitals pointing towards the vacancy. These orbitals cannot be obtained by wannierizing the three visible defect levels in the gap alone, as the nodeless combination which lies within the valence band is also needed; therefore, the entire valence manifold plus the defect levels in the gap must be wannierized as a whole.

The defect Wannier functions obtained for $\mathrm{v}^{2+}$, $\mathrm{v}^0$, and $\mathrm{v}^{1-}$ in their relaxed configurations (Figs.~\ref{subfig:wan-2+}--\ref{subfig:wan-1-}) also support Watkins' model. For the doubly positive vacancy, the defect levels in the gap are empty and therefore not included; the result is a single MLWF corresponding to the symmetric $s$-type nodeless combination of the four $sp^3$ orbitals. The neutral vacancy includes one lowered defect level only, and produces the two bonding orbitals between the pairs of nearest neighbours of the defect centre. The singly negative vacancy includes two defect levels in the gap (for the majority spin), producing one bonding orbital (between the pair of ions with the shortest bond length) and two $sp^3$ orbitals.

Finally, we show the split vacancy configuration obtained for $\mathrm{v}^{2-}$ in Figs.~\ref{subfig:label-split}--\ref{subfig:wan-2-}. In this case, one of the neighbours (labelled c) moves halfway along the line connecting it with the vacancy site, thus placing itself at the centre of an octahedron made up of two pairs of second-nearest neighbours (labelled a and b). This octahedron is approximately regular (having as faces equilateral triangles); however, there is a small distortion due to the difference in distance between a--a/b--b pairs and a--b pairs. This distortion is quantified by the angle $\alpha$ between any three ions a, c, b; for a regular octahedron, $\alpha$ is a right angle, and so the distances a--a, b--b, a--b are identical. In general, the split configuration can be specified with only two parameters: the distortion angle $\alpha$ and the distance a--c between the central ion and any of the ions forming the octahedral cage.

The wannierization of the occupied manifold produces six defect Wannier functions, each one a bond between the central ion and one of its neighbours. This suggests that the c ion is forming six $sp^3d^2$ orbitals which then bond to the dangling $sp^3$ orbitals of the a, b ions. The inwards relaxation of the ionic positions also shortens the bonds and reduces the distortion of the octahedron. This configuration is favoured since these bonds are shorter than the ones obtained by dimerization in the $\mathrm{C_{2v}}$ arrangement predicted by Watkins' model, and so are closer in length to the bulk silicon bond. However, only in the case of the doubly negative vacancy are there enough electrons to fully occupy the six orbitals.


\begin{thebibliography}{39}%
\makeatletter
\providecommand \@ifxundefined [1]{%
 \@ifx{#1\undefined}
}%
\providecommand \@ifnum [1]{%
 \ifnum #1\expandafter \@firstoftwo
 \else \expandafter \@secondoftwo
 \fi
}%
\providecommand \@ifx [1]{%
 \ifx #1\expandafter \@firstoftwo
 \else \expandafter \@secondoftwo
 \fi
}%
\providecommand \natexlab [1]{#1}%
\providecommand \enquote  [1]{``#1''}%
\providecommand \bibnamefont  [1]{#1}%
\providecommand \bibfnamefont [1]{#1}%
\providecommand \citenamefont [1]{#1}%
\providecommand \href@noop [0]{\@secondoftwo}%
\providecommand \href [0]{\begingroup \@sanitize@url \@href}%
\providecommand \@href[1]{\@@startlink{#1}\@@href}%
\providecommand \@@href[1]{\endgroup#1\@@endlink}%
\providecommand \@sanitize@url [0]{\catcode `\\12\catcode `\$12\catcode
  `\&12\catcode `\#12\catcode `\^12\catcode `\_12\catcode `\%12\relax}%
\providecommand \@@startlink[1]{}%
\providecommand \@@endlink[0]{}%
\providecommand \url  [0]{\begingroup\@sanitize@url \@url }%
\providecommand \@url [1]{\endgroup\@href {#1}{\urlprefix }}%
\providecommand \urlprefix  [0]{URL }%
\providecommand \Eprint [0]{\href }%
\providecommand \doibase [0]{http://dx.doi.org/}%
\providecommand \selectlanguage [0]{\@gobble}%
\providecommand \bibinfo  [0]{\@secondoftwo}%
\providecommand \bibfield  [0]{\@secondoftwo}%
\providecommand \translation [1]{[#1]}%
\providecommand \BibitemOpen [0]{}%
\providecommand \bibitemStop [0]{}%
\providecommand \bibitemNoStop [0]{.\EOS\space}%
\providecommand \EOS [0]{\spacefactor3000\relax}%
\providecommand \BibitemShut  [1]{\csname bibitem#1\endcsname}%
\let\auto@bib@innerbib\@empty
\bibitem [{\citenamefont {Van~de Walle}\ and\ \citenamefont
  {Neugebauer}(2004)}]{defect-rev}%
  \BibitemOpen
  \bibfield  {author} {\bibinfo {author} {\bibfnamefont {C.~G.}\ \bibnamefont
  {Van~de Walle}}\ and\ \bibinfo {author} {\bibfnamefont {J.}~\bibnamefont
  {Neugebauer}},\ }\href@noop {} {\bibfield  {journal} {\bibinfo  {journal} {J.
  Appl. Phys.}\ }\textbf {\bibinfo {volume} {95}},\ \bibinfo {pages} {3851}
  (\bibinfo {year} {2004})}\BibitemShut {NoStop}%
\bibitem [{\citenamefont {Kohn}\ and\ \citenamefont {Sham}(1965)}]{ks}%
  \BibitemOpen
  \bibfield  {author} {\bibinfo {author} {\bibfnamefont {W.}~\bibnamefont
  {Kohn}}\ and\ \bibinfo {author} {\bibfnamefont {L.~J.}\ \bibnamefont
  {Sham}},\ }\href@noop {} {\bibfield  {journal} {\bibinfo  {journal} {Phys.
  Rev.}\ }\textbf {\bibinfo {volume} {140}},\ \bibinfo {pages} {A1133}
  (\bibinfo {year} {1965})}\BibitemShut {NoStop}%
\bibitem [{\citenamefont {Ganchenkova}\ \emph {et~al.}(2009)\citenamefont
  {Ganchenkova}, \citenamefont {Oikkonen}, \citenamefont {Borodin},
  \citenamefont {Nicolaysen},\ and\ \citenamefont {Nieminen}}]{niem-multisymm}%
  \BibitemOpen
  \bibfield  {author} {\bibinfo {author} {\bibfnamefont {M.~G.}\ \bibnamefont
  {Ganchenkova}}, \bibinfo {author} {\bibfnamefont {L.~E.}\ \bibnamefont
  {Oikkonen}}, \bibinfo {author} {\bibfnamefont {V.~A.}\ \bibnamefont
  {Borodin}}, \bibinfo {author} {\bibfnamefont {S.}~\bibnamefont {Nicolaysen}},
  \ and\ \bibinfo {author} {\bibfnamefont {R.~M.}\ \bibnamefont {Nieminen}},\
  }\href@noop {} {\bibfield  {journal} {\bibinfo  {journal} {Mater. Sci. Eng.
  B}\ }\textbf {\bibinfo {volume} {159-160}},\ \bibinfo {pages} {107} (\bibinfo
  {year} {2009})}\BibitemShut {NoStop}%
\bibitem [{\citenamefont {Probert}\ and\ \citenamefont
  {Payne}(2003)}]{probert}%
  \BibitemOpen
  \bibfield  {author} {\bibinfo {author} {\bibfnamefont {M.~I.~J.}\
  \bibnamefont {Probert}}\ and\ \bibinfo {author} {\bibfnamefont {M.~C.}\
  \bibnamefont {Payne}},\ }\href@noop {} {\bibfield  {journal} {\bibinfo
  {journal} {Phys. Rev. B}\ }\textbf {\bibinfo {volume} {67}},\ \bibinfo
  {pages} {075204} (\bibinfo {year} {2003})}\BibitemShut {NoStop}%
\bibitem [{Note1()}]{Note1}%
  \BibitemOpen
  \bibinfo {note} {For a review, see Probert and Payne\cite {probert} and Puska
  {\protect \em et al.}\cite {niem-vac}.}\BibitemShut {Stop}%
\bibitem [{\citenamefont {Puska}\ \emph {et~al.}(1998)\citenamefont {Puska},
  \citenamefont {P\"oykk\"o}, \citenamefont {Pesola},\ and\ \citenamefont
  {Nieminen}}]{niem-vac}%
  \BibitemOpen
  \bibfield  {author} {\bibinfo {author} {\bibfnamefont {M.~J.}\ \bibnamefont
  {Puska}}, \bibinfo {author} {\bibfnamefont {S.}~\bibnamefont {P\"oykk\"o}},
  \bibinfo {author} {\bibfnamefont {M.}~\bibnamefont {Pesola}}, \ and\ \bibinfo
  {author} {\bibfnamefont {R.~M.}\ \bibnamefont {Nieminen}},\ }\href {\doibase
  10.1103/PhysRevB.58.1318} {\bibfield  {journal} {\bibinfo  {journal} {Phys.
  Rev. B}\ }\textbf {\bibinfo {volume} {58}},\ \bibinfo {pages} {1318}
  (\bibinfo {year} {1998})}\BibitemShut {NoStop}%
\bibitem [{\citenamefont {Wright}(2006)}]{wright-vac}%
  \BibitemOpen
  \bibfield  {author} {\bibinfo {author} {\bibfnamefont {A.~F.}\ \bibnamefont
  {Wright}},\ }\href {\doibase 10.1103/PhysRevB.74.165116} {\bibfield
  {journal} {\bibinfo  {journal} {Phys. Rev. B}\ }\textbf {\bibinfo {volume}
  {74}},\ \bibinfo {pages} {165116} (\bibinfo {year} {2006})}\BibitemShut
  {NoStop}%
\bibitem [{\citenamefont {Marzari}\ and\ \citenamefont
  {Vanderbilt}(1997)}]{mlwf}%
  \BibitemOpen
  \bibfield  {author} {\bibinfo {author} {\bibfnamefont {N.}~\bibnamefont
  {Marzari}}\ and\ \bibinfo {author} {\bibfnamefont {D.}~\bibnamefont
  {Vanderbilt}},\ }\href@noop {} {\bibfield  {journal} {\bibinfo  {journal}
  {Phys. Rev. B}\ }\textbf {\bibinfo {volume} {56}},\ \bibinfo {pages} {12847}
  (\bibinfo {year} {1997})}\BibitemShut {NoStop}%
\bibitem [{\citenamefont {Watkins}(1986)}]{watkins-model}%
  \BibitemOpen
  \bibfield  {author} {\bibinfo {author} {\bibfnamefont {G.~D.}\ \bibnamefont
  {Watkins}},\ }in\ \href@noop {} {\emph {\bibinfo {booktitle} {Deep Centres in
  Semiconductors}}},\ \bibinfo {editor} {edited by\ \bibinfo {editor}
  {\bibfnamefont {S.~T.}\ \bibnamefont {Pantelides}}}\ (\bibinfo  {publisher}
  {Gordon and Breach},\ \bibinfo {address} {New York},\ \bibinfo {year}
  {1986})\ p.\ \bibinfo {pages} {147}\BibitemShut {NoStop}%
\bibitem [{\citenamefont {Baraff}\ \emph {et~al.}(1980)\citenamefont {Baraff},
  \citenamefont {Kane},\ and\ \citenamefont {Schl\"{u}ter}}]{negU-baraff}%
  \BibitemOpen
  \bibfield  {author} {\bibinfo {author} {\bibfnamefont {G.~A.}\ \bibnamefont
  {Baraff}}, \bibinfo {author} {\bibfnamefont {E.~O.}\ \bibnamefont {Kane}}, \
  and\ \bibinfo {author} {\bibfnamefont {M.}~\bibnamefont {Schl\"{u}ter}},\
  }\href@noop {} {\bibfield  {journal} {\bibinfo  {journal} {Phys. Rev. B}\
  }\textbf {\bibinfo {volume} {21}},\ \bibinfo {pages} {5662} (\bibinfo {year}
  {1980})}\BibitemShut {NoStop}%
\bibitem [{\citenamefont {Watkins}(1983)}]{vac}%
  \BibitemOpen
  \bibfield  {author} {\bibinfo {author} {\bibfnamefont {G.~D.}\ \bibnamefont
  {Watkins}},\ }\href@noop {} {\bibfield  {journal} {\bibinfo  {journal}
  {Physica B+C}\ }\textbf {\bibinfo {volume} {117B-118B}},\ \bibinfo {pages}
  {9} (\bibinfo {year} {1983})}\BibitemShut {NoStop}%
\bibitem [{\citenamefont {Zhang}\ and\ \citenamefont {Northrup}(1991)}]{zhang}%
  \BibitemOpen
  \bibfield  {author} {\bibinfo {author} {\bibfnamefont {S.~B.}\ \bibnamefont
  {Zhang}}\ and\ \bibinfo {author} {\bibfnamefont {J.~E.}\ \bibnamefont
  {Northrup}},\ }\href@noop {} {\bibfield  {journal} {\bibinfo  {journal}
  {Phys. Rev. Lett.}\ }\textbf {\bibinfo {volume} {67}},\ \bibinfo {pages}
  {2339} (\bibinfo {year} {1991})}\BibitemShut {NoStop}%
\bibitem [{\citenamefont {Clark}\ \emph {et~al.}(2005)\citenamefont {Clark},
  \citenamefont {Segall}, \citenamefont {Pickard}, \citenamefont {Hasnip},
  \citenamefont {Probert}, \citenamefont {Refson},\ and\ \citenamefont
  {Payne}}]{castep}%
  \BibitemOpen
  \bibfield  {author} {\bibinfo {author} {\bibfnamefont {S.~J.}\ \bibnamefont
  {Clark}}, \bibinfo {author} {\bibfnamefont {M.~D.}\ \bibnamefont {Segall}},
  \bibinfo {author} {\bibfnamefont {C.~J.}\ \bibnamefont {Pickard}}, \bibinfo
  {author} {\bibfnamefont {P.~J.}\ \bibnamefont {Hasnip}}, \bibinfo {author}
  {\bibfnamefont {M.~I.~J.}\ \bibnamefont {Probert}}, \bibinfo {author}
  {\bibfnamefont {K.}~\bibnamefont {Refson}}, \ and\ \bibinfo {author}
  {\bibfnamefont {M.~C.}\ \bibnamefont {Payne}},\ }\href@noop {} {\bibfield
  {journal} {\bibinfo  {journal} {Z. Kristallogr.}\ }\textbf {\bibinfo {volume}
  {220}},\ \bibinfo {pages} {567} (\bibinfo {year} {2005})}\BibitemShut
  {NoStop}%
\bibitem [{\citenamefont {Ceperley}\ and\ \citenamefont {Alder}(1980)}]{qmc1}%
  \BibitemOpen
  \bibfield  {author} {\bibinfo {author} {\bibfnamefont {D.~M.}\ \bibnamefont
  {Ceperley}}\ and\ \bibinfo {author} {\bibfnamefont {B.~J.}\ \bibnamefont
  {Alder}},\ }\href@noop {} {\bibfield  {journal} {\bibinfo  {journal} {Phys.
  Rev. Lett.}\ }\textbf {\bibinfo {volume} {45}},\ \bibinfo {pages} {566}
  (\bibinfo {year} {1980})}\BibitemShut {NoStop}%
\bibitem [{\citenamefont {Vanderbilt}(1990)}]{ultra-pseudo}%
  \BibitemOpen
  \bibfield  {author} {\bibinfo {author} {\bibfnamefont {D.}~\bibnamefont
  {Vanderbilt}},\ }\href@noop {} {\bibfield  {journal} {\bibinfo  {journal}
  {Phys. Rev. B}\ }\textbf {\bibinfo {volume} {41}},\ \bibinfo {pages} {7892}
  (\bibinfo {year} {1990})}\BibitemShut {NoStop}%
\bibitem [{\citenamefont {Hamann}\ \emph {et~al.}(1979)\citenamefont {Hamann},
  \citenamefont {Schl\"{u}ter},\ and\ \citenamefont {Chiang}}]{norm-pseudo}%
  \BibitemOpen
  \bibfield  {author} {\bibinfo {author} {\bibfnamefont {D.~R.}\ \bibnamefont
  {Hamann}}, \bibinfo {author} {\bibfnamefont {M.}~\bibnamefont
  {Schl\"{u}ter}}, \ and\ \bibinfo {author} {\bibfnamefont {C.}~\bibnamefont
  {Chiang}},\ }\href@noop {} {\bibfield  {journal} {\bibinfo  {journal} {Phys.
  Rev. Lett.}\ }\textbf {\bibinfo {volume} {43}},\ \bibinfo {pages} {1494}
  (\bibinfo {year} {1979})}\BibitemShut {NoStop}%
\bibitem [{\citenamefont {Skylaris}\ \emph {et~al.}(2005)\citenamefont
  {Skylaris}, \citenamefont {Haynes}, \citenamefont {Mostofi},\ and\
  \citenamefont {Payne}}]{onetep1}%
  \BibitemOpen
  \bibfield  {author} {\bibinfo {author} {\bibfnamefont {C.-K.}\ \bibnamefont
  {Skylaris}}, \bibinfo {author} {\bibfnamefont {P.~D.}\ \bibnamefont
  {Haynes}}, \bibinfo {author} {\bibfnamefont {A.~A.}\ \bibnamefont {Mostofi}},
  \ and\ \bibinfo {author} {\bibfnamefont {M.~C.}\ \bibnamefont {Payne}},\
  }\href@noop {} {\bibfield  {journal} {\bibinfo  {journal} {J. Chem. Phys.}\
  }\textbf {\bibinfo {volume} {122}},\ \bibinfo {pages} {084119} (\bibinfo
  {year} {2005})}\BibitemShut {NoStop}%
\bibitem [{\citenamefont {Hine}\ \emph
  {et~al.}(2009{\natexlab{a}})\citenamefont {Hine}, \citenamefont {Haynes},
  \citenamefont {Mostofi}, \citenamefont {Skylaris},\ and\ \citenamefont
  {Payne}}]{onetep-nick}%
  \BibitemOpen
  \bibfield  {author} {\bibinfo {author} {\bibfnamefont {N.~D.~M.}\
  \bibnamefont {Hine}}, \bibinfo {author} {\bibfnamefont {P.~D.}\ \bibnamefont
  {Haynes}}, \bibinfo {author} {\bibfnamefont {A.~A.}\ \bibnamefont {Mostofi}},
  \bibinfo {author} {\bibfnamefont {C.-K.}\ \bibnamefont {Skylaris}}, \ and\
  \bibinfo {author} {\bibfnamefont {M.~C.}\ \bibnamefont {Payne}},\ }\href@noop
  {} {\bibfield  {journal} {\bibinfo  {journal} {Comput. Phys. Commun.}\
  }\textbf {\bibinfo {volume} {180}},\ \bibinfo {pages} {1041} (\bibinfo {year}
  {2009}{\natexlab{a}})}\BibitemShut {NoStop}%
\bibitem [{\citenamefont {Monkhorst}\ and\ \citenamefont
  {Pack}(1976)}]{mp_grid}%
  \BibitemOpen
  \bibfield  {author} {\bibinfo {author} {\bibfnamefont {H.~J.}\ \bibnamefont
  {Monkhorst}}\ and\ \bibinfo {author} {\bibfnamefont {J.~D.}\ \bibnamefont
  {Pack}},\ }\href {\doibase 10.1103/PhysRevB.13.5188} {\bibfield  {journal}
  {\bibinfo  {journal} {Phys. Rev. B}\ }\textbf {\bibinfo {volume} {13}},\
  \bibinfo {pages} {5188} (\bibinfo {year} {1976})}\BibitemShut {NoStop}%
\bibitem [{\citenamefont {Persson}\ \emph {et~al.}(2005)\citenamefont
  {Persson}, \citenamefont {Zhao}, \citenamefont {Lany},\ and\ \citenamefont
  {Zunger}}]{vbm_hole}%
  \BibitemOpen
  \bibfield  {author} {\bibinfo {author} {\bibfnamefont {C.}~\bibnamefont
  {Persson}}, \bibinfo {author} {\bibfnamefont {Y.-J.}\ \bibnamefont {Zhao}},
  \bibinfo {author} {\bibfnamefont {S.}~\bibnamefont {Lany}}, \ and\ \bibinfo
  {author} {\bibfnamefont {A.}~\bibnamefont {Zunger}},\ }\href {\doibase
  10.1103/PhysRevB.72.035211} {\bibfield  {journal} {\bibinfo  {journal} {Phys.
  Rev. B}\ }\textbf {\bibinfo {volume} {72}},\ \bibinfo {pages} {035211}
  (\bibinfo {year} {2005})}\BibitemShut {NoStop}%
\bibitem [{\citenamefont {Lany}\ and\ \citenamefont {Zunger}(2008)}]{def_corr}%
  \BibitemOpen
  \bibfield  {author} {\bibinfo {author} {\bibfnamefont {S.}~\bibnamefont
  {Lany}}\ and\ \bibinfo {author} {\bibfnamefont {A.}~\bibnamefont {Zunger}},\
  }\href {\doibase 10.1103/PhysRevB.78.235104} {\bibfield  {journal} {\bibinfo
  {journal} {Phys. Rev. B}\ }\textbf {\bibinfo {volume} {78}},\ \bibinfo
  {pages} {235104} (\bibinfo {year} {2008})}\BibitemShut {NoStop}%
\bibitem [{\citenamefont {P\"oykk\"o}\ \emph {et~al.}(1996)\citenamefont
  {P\"oykk\"o}, \citenamefont {Puska},\ and\ \citenamefont
  {Nieminen}}]{niem-gaas}%
  \BibitemOpen
  \bibfield  {author} {\bibinfo {author} {\bibfnamefont {S.}~\bibnamefont
  {P\"oykk\"o}}, \bibinfo {author} {\bibfnamefont {M.~J.}\ \bibnamefont
  {Puska}}, \ and\ \bibinfo {author} {\bibfnamefont {R.~M.}\ \bibnamefont
  {Nieminen}},\ }\href {\doibase 10.1103/PhysRevB.53.3813} {\bibfield
  {journal} {\bibinfo  {journal} {Phys. Rev. B}\ }\textbf {\bibinfo {volume}
  {53}},\ \bibinfo {pages} {3813} (\bibinfo {year} {1996})}\BibitemShut
  {NoStop}%
\bibitem [{Note2()}]{Note2}%
  \BibitemOpen
  \bibinfo {note} {Only the valence bands have been taken into account for the
  wannierization.}\BibitemShut {Stop}%
\bibitem [{\citenamefont {Mostofi}\ \emph {et~al.}(2008)\citenamefont
  {Mostofi}, \citenamefont {Yates}, \citenamefont {Lee}, \citenamefont {Souza},
  \citenamefont {Vanderbilt},\ and\ \citenamefont {Marzari}}]{wan90}%
  \BibitemOpen
  \bibfield  {author} {\bibinfo {author} {\bibfnamefont {A.~A.}\ \bibnamefont
  {Mostofi}}, \bibinfo {author} {\bibfnamefont {J.~R.}\ \bibnamefont {Yates}},
  \bibinfo {author} {\bibfnamefont {Y.-S.}\ \bibnamefont {Lee}}, \bibinfo
  {author} {\bibfnamefont {I.}~\bibnamefont {Souza}}, \bibinfo {author}
  {\bibfnamefont {D.}~\bibnamefont {Vanderbilt}}, \ and\ \bibinfo {author}
  {\bibfnamefont {N.}~\bibnamefont {Marzari}},\ }\href@noop {} {\bibfield
  {journal} {\bibinfo  {journal} {Comput. Phys. Commun.}\ }\textbf {\bibinfo
  {volume} {178}},\ \bibinfo {pages} {685} (\bibinfo {year}
  {2008})}\BibitemShut {NoStop}%
\bibitem [{Note3()}]{Note3}%
  \BibitemOpen
  \bibinfo {note} {J. Neugebauer, private communication.}\BibitemShut {Stop}%
\bibitem [{Note4()}]{Note4}%
  \BibitemOpen
  \bibinfo {note} {The unrelaxed defect formation energy has not been
  calculated with {\protect \sc onetep}, since the density matrix method does
  not allow for a degenerate ground state without an explicit
  (non-linear-scaling) diagonalization.}\BibitemShut {Stop}%
\bibitem [{Note5()}]{Note5}%
  \BibitemOpen
  \bibinfo {note} {The results given use the Voronoi cell method for
  determining the potential alignment; this is to ensure consistency among all
  results, since for the 1000-atom {\protect \sc onetep} calculation only the
  Voronoi cell method is appropriate. For all comparisons of {\protect \sc
  castep} calculations between the Voronoi cell and MLWFs methods the agreement
  was to within 0.01~eV (calculations not shown).}\BibitemShut {Stop}%
\bibitem [{\citenamefont {Watkins}\ and\ \citenamefont {Troxell}(1980)}]{negU}%
  \BibitemOpen
  \bibfield  {author} {\bibinfo {author} {\bibfnamefont {G.~D.}\ \bibnamefont
  {Watkins}}\ and\ \bibinfo {author} {\bibfnamefont {J.~R.}\ \bibnamefont
  {Troxell}},\ }\href@noop {} {\bibfield  {journal} {\bibinfo  {journal} {Phys.
  Rev. Lett.}\ }\textbf {\bibinfo {volume} {44}},\ \bibinfo {pages} {593}
  (\bibinfo {year} {1980})}\BibitemShut {NoStop}%
\bibitem [{\citenamefont {Makov}\ and\ \citenamefont
  {Payne}(1995)}]{interactions}%
  \BibitemOpen
  \bibfield  {author} {\bibinfo {author} {\bibfnamefont {G.}~\bibnamefont
  {Makov}}\ and\ \bibinfo {author} {\bibfnamefont {M.~C.}\ \bibnamefont
  {Payne}},\ }\href@noop {} {\bibfield  {journal} {\bibinfo  {journal} {Phys.
  Rev. B}\ }\textbf {\bibinfo {volume} {51}},\ \bibinfo {pages} {4014}
  (\bibinfo {year} {1995})}\BibitemShut {NoStop}%
\bibitem [{\citenamefont {Castleton}\ \emph {et~al.}(2006)\citenamefont
  {Castleton}, \citenamefont {H\"{o}glund},\ and\ \citenamefont
  {Mirbt}}]{gap-prob}%
  \BibitemOpen
  \bibfield  {author} {\bibinfo {author} {\bibfnamefont {C.~W.~M.}\
  \bibnamefont {Castleton}}, \bibinfo {author} {\bibfnamefont {A.}~\bibnamefont
  {H\"{o}glund}}, \ and\ \bibinfo {author} {\bibfnamefont {S.}~\bibnamefont
  {Mirbt}},\ }\href@noop {} {\bibfield  {journal} {\bibinfo  {journal} {Phys.
  Rev. B}\ }\textbf {\bibinfo {volume} {73}},\ \bibinfo {pages} {035215}
  (\bibinfo {year} {2006})}\BibitemShut {NoStop}%
\bibitem [{\citenamefont {Alkauskas}\ \emph {et~al.}(2008)\citenamefont
  {Alkauskas}, \citenamefont {Broqvist},\ and\ \citenamefont
  {Pasquarello}}]{hybrid2}%
  \BibitemOpen
  \bibfield  {author} {\bibinfo {author} {\bibfnamefont {A.}~\bibnamefont
  {Alkauskas}}, \bibinfo {author} {\bibfnamefont {P.}~\bibnamefont {Broqvist}},
  \ and\ \bibinfo {author} {\bibfnamefont {A.}~\bibnamefont {Pasquarello}},\
  }\href@noop {} {\bibfield  {journal} {\bibinfo  {journal} {Phys. Rev. Lett.}\
  }\textbf {\bibinfo {volume} {101}},\ \bibinfo {pages} {046405} (\bibinfo
  {year} {2008})}\BibitemShut {NoStop}%
\bibitem [{\citenamefont {Freysoldt}\ \emph {et~al.}(2009)\citenamefont
  {Freysoldt}, \citenamefont {Neugebauer},\ and\ \citenamefont {Van~de
  Walle}}]{mp-over2}%
  \BibitemOpen
  \bibfield  {author} {\bibinfo {author} {\bibfnamefont {C.}~\bibnamefont
  {Freysoldt}}, \bibinfo {author} {\bibfnamefont {J.}~\bibnamefont
  {Neugebauer}}, \ and\ \bibinfo {author} {\bibfnamefont {C.~G.}\ \bibnamefont
  {Van~de Walle}},\ }\href {\doibase 10.1103/PhysRevLett.102.016402} {\bibfield
   {journal} {\bibinfo  {journal} {Phys. Rev. Lett.}\ }\textbf {\bibinfo
  {volume} {102}},\ \bibinfo {pages} {016402} (\bibinfo {year}
  {2009})}\BibitemShut {NoStop}%
\bibitem [{\citenamefont {Hine}\ \emph
  {et~al.}(2009{\natexlab{b}})\citenamefont {Hine}, \citenamefont {Frensch},
  \citenamefont {Foulkes},\ and\ \citenamefont
  {Finnis}}]{nick-supercell_shape}%
  \BibitemOpen
  \bibfield  {author} {\bibinfo {author} {\bibfnamefont {N.~D.~M.}\
  \bibnamefont {Hine}}, \bibinfo {author} {\bibfnamefont {K.}~\bibnamefont
  {Frensch}}, \bibinfo {author} {\bibfnamefont {W.~M.~C.}\ \bibnamefont
  {Foulkes}}, \ and\ \bibinfo {author} {\bibfnamefont {M.~W.}\ \bibnamefont
  {Finnis}},\ }\href {\doibase 10.1103/PhysRevB.79.024112} {\bibfield
  {journal} {\bibinfo  {journal} {Phys. Rev. B}\ }\textbf {\bibinfo {volume}
  {79}},\ \bibinfo {pages} {024112} (\bibinfo {year}
  {2009}{\natexlab{b}})}\BibitemShut {NoStop}%
\bibitem [{\citenamefont {Lento}\ and\ \citenamefont
  {Nieminen}(2003)}]{sx_lda-vac}%
  \BibitemOpen
  \bibfield  {author} {\bibinfo {author} {\bibfnamefont {J.}~\bibnamefont
  {Lento}}\ and\ \bibinfo {author} {\bibfnamefont {R.~M.}\ \bibnamefont
  {Nieminen}},\ }\href@noop {} {\bibfield  {journal} {\bibinfo  {journal} {J.
  Phys.: Condens. Matter}\ }\textbf {\bibinfo {volume} {15}},\ \bibinfo {pages}
  {4387} (\bibinfo {year} {2003})}\BibitemShut {NoStop}%
\bibitem [{\citenamefont {Souza}\ \emph {et~al.}(2001)\citenamefont {Souza},
  \citenamefont {Marzari},\ and\ \citenamefont {Vanderbilt}}]{wan-ent}%
  \BibitemOpen
  \bibfield  {author} {\bibinfo {author} {\bibfnamefont {I.}~\bibnamefont
  {Souza}}, \bibinfo {author} {\bibfnamefont {N.}~\bibnamefont {Marzari}}, \
  and\ \bibinfo {author} {\bibfnamefont {D.}~\bibnamefont {Vanderbilt}},\
  }\href@noop {} {\bibfield  {journal} {\bibinfo  {journal} {Phys. Rev. B}\
  }\textbf {\bibinfo {volume} {65}},\ \bibinfo {pages} {035109} (\bibinfo
  {year} {2001})}\BibitemShut {NoStop}%
\bibitem [{\citenamefont {Watkins}(1976)}]{jt-exp}%
  \BibitemOpen
  \bibfield  {author} {\bibinfo {author} {\bibfnamefont {G.~D.}\ \bibnamefont
  {Watkins}},\ }in\ \href@noop {} {\emph {\bibinfo {booktitle} {Defects and
  Their Structure in Non-metallic Solids}}},\ \bibinfo {editor} {edited by\
  \bibinfo {editor} {\bibfnamefont {B.}~\bibnamefont {Henderson}}\ and\
  \bibinfo {editor} {\bibfnamefont {A.~E.}\ \bibnamefont {Hughes}}}\ (\bibinfo
  {publisher} {Plenum},\ \bibinfo {address} {New York},\ \bibinfo {year}
  {1976})\ p.\ \bibinfo {pages} {203}\BibitemShut {NoStop}%
\bibitem [{\citenamefont {Sprenger}\ \emph {et~al.}(1983)\citenamefont
  {Sprenger}, \citenamefont {Muller},\ and\ \citenamefont
  {Ammerlaan}}]{endor1}%
  \BibitemOpen
  \bibfield  {author} {\bibinfo {author} {\bibfnamefont {M.}~\bibnamefont
  {Sprenger}}, \bibinfo {author} {\bibfnamefont {S.~H.}\ \bibnamefont
  {Muller}}, \ and\ \bibinfo {author} {\bibfnamefont {C.~A.~J.}\ \bibnamefont
  {Ammerlaan}},\ }\href@noop {} {\bibfield  {journal} {\bibinfo  {journal}
  {Physica B+C}\ }\textbf {\bibinfo {volume} {116B}},\ \bibinfo {pages} {224}
  (\bibinfo {year} {1983})}\BibitemShut {NoStop}%
\bibitem [{\citenamefont {Sprenger}\ \emph {et~al.}(1987)\citenamefont
  {Sprenger}, \citenamefont {Muller}, \citenamefont {Sieverts},\ and\
  \citenamefont {Ammerlaan}}]{endor2}%
  \BibitemOpen
  \bibfield  {author} {\bibinfo {author} {\bibfnamefont {M.}~\bibnamefont
  {Sprenger}}, \bibinfo {author} {\bibfnamefont {S.~H.}\ \bibnamefont
  {Muller}}, \bibinfo {author} {\bibfnamefont {E.~G.}\ \bibnamefont
  {Sieverts}}, \ and\ \bibinfo {author} {\bibfnamefont {C.~A.~J.}\ \bibnamefont
  {Ammerlaan}},\ }\href {\doibase 10.1103/PhysRevB.35.1566} {\bibfield
  {journal} {\bibinfo  {journal} {Phys. Rev. B}\ }\textbf {\bibinfo {volume}
  {35}},\ \bibinfo {pages} {1566} (\bibinfo {year} {1987})}\BibitemShut
  {NoStop}%
\bibitem [{\citenamefont {Jahn}\ and\ \citenamefont
  {Teller}(1937)}]{jt-theorem}%
  \BibitemOpen
  \bibfield  {author} {\bibinfo {author} {\bibfnamefont {H.~A.}\ \bibnamefont
  {Jahn}}\ and\ \bibinfo {author} {\bibfnamefont {E.}~\bibnamefont {Teller}},\
  }\href@noop {} {\bibfield  {journal} {\bibinfo  {journal} {Proc. R. Soc.
  Lond. A}\ }\textbf {\bibinfo {volume} {161}},\ \bibinfo {pages} {220}
  (\bibinfo {year} {1937})}\BibitemShut {NoStop}%
\end{thebibliography}
\end{document}